%% file: master-lncs-lata2021.tex
\documentclass[final,UKenglish,runningheads,envcountsect,envcountsame,final]{llncs}

\usepackage{wrapfig}
\usepackage{amssymb}
\usepackage{amsmath}
\usepackage{stmaryrd}
\usepackage{mathtools}
\usepackage{amsmath}
\usepackage{xspace}
\usepackage{tikz-cd}
\usepackage{multicol}
\usepackage{dsfont}
\usepackage{mathrsfs}

\usepackage[notref,notcite]{showkeys}
\usepackage[noadjust]{cite}

\newcommand{\takeout}[1]{\empty}
\newcommand{\dun}{\mathbin{\dot\cup}}
\newcommand{\Int}{\mathds{Z}}

\tikzset{shiftarr/.style={
        rounded corners,%
        to path={--([#1]\tikztostart.center)
                     -- ([#1]\tikztotarget.center) \tikztonodes
                     -- (\tikztotarget)},
}}

\include{unicode}

\usepackage{etex,etoolbox}


\usepackage{ifdraft}
\ifdraft{
\usepackage[nohead,nofoot,
            headsep = 1cm,
            footskip = 1cm,
            text={12.2cm,19.3cm},
            paperwidth=16.2cm,
            paperheight=23.3cm,
            marginparsep=1mm,
            marginparwidth=1.88cm,
            footskip=1cm,
            centering
            ]{geometry}
          }{}

\usepackage{hyperref}
\hypersetup{
  hidelinks,final,bookmarks
}

\usepackage{braket}
\renewcommand{\Set}{\ensuremath{\cat{Set}}\xspace}
\usepackage{seqsplit}
\usepackage{xstring}
\newcommand{\defaultshowkeysformat}[1]{%
\StrSubstitute{#1}{ }{\textvisiblespace}[\TEMP]%
\parbox[t]{\marginparwidth}{\raggedright\normalfont\small\ttfamily\(\{\){\color{red!50!black}\expandafter\seqsplit\expandafter{\TEMP}}\(\}\)}%
}

\renewcommand*\showkeyslabelformat[1]{%
\noexpandarg%
\defaultshowkeysformat{#1}%
}

\usepackage[inline]{enumitem}
\setlist[enumerate,1]{label=(\arabic*),font=\normalfont,align=left,leftmargin=0pt,labelindent=0pt,listparindent=\parindent,labelwidth=0pt,itemindent=!,topsep=3pt,parsep=0pt,itemsep=3pt,start=1}
\setlist[enumerate,2]{label=(\alph*),font=\normalfont,labelindent=*,leftmargin=*,start=1}
\setlist[itemize]{labelindent=*,leftmargin=*,topsep=5pt,itemsep=3pt}
\setlist[description]{labelindent=*,leftmargin=*,itemindent=-1 em}

\makeatletter
\renewcommand\section{\@startsection{section}{1}{\z@}%
  {-18\p@ \@plus -4\p@ \@minus -4\p@}%
  {12\p@ \@plus 4\p@ \@minus 4\p@}%
  {\normalfont\large\bfseries\boldmath
    \rightskip=\z@ \@plus 8em\pretolerance=10000 }}
\renewcommand\subsection{\@startsection{subsection}{2}{\z@}%
 {-6\p@ \@plus -2\p@ \@minus -2\p@}%
 {-0.5em \@plus -0.22em \@minus -0.1em}%
 {\normalfont\normalsize\bfseries\boldmath}}                      

\newcommand\mysubsec{\@startsection{paragraph}{4}{\z@}%
  {-6\p@ \@plus -4\p@ \@minus -4\p@}%
  {-0.5em \@plus -0.22em \@minus -0.1em}%
  {\normalfont\normalsize\bfseries}}
\makeatother

\usepackage[footnote,marginclue,nomargin]{fixme}

\spnewtheorem{assumptions}[theorem]{Assumptions}{\bfseries}{\rmfamily}
\spnewtheorem{notation}[theorem]{Notation}{\bfseries}{\rmfamily}
\spnewtheorem{observation}[theorem]{Observation}{\bfseries}{\rmfamily}
\spnewtheorem{defn}[theorem]{Definition}{\bfseries}{\rmfamily}
\spnewtheorem{expl}[theorem]{Example}{\bfseries}{\rmfamily}
\spnewtheorem{rem}[theorem]{Remark}{\bfseries}{\rmfamily}
\spnewtheorem{fact}[theorem]{Fact}{\bfseries}{\rmfamily}
\spnewtheorem{construction}[theorem]{Construction}{\bfseries}{\rmfamily}
\spnewtheorem{examples}[theorem]{Examples}{\bfseries}{\rmfamily}
\spnewtheorem*{bvthm}{Basic Variety Theorem}{\normalfont\bfseries}{\itshape}

\numberwithin{equation}{section}

\begin{document}
\FXRegisterAuthor{fb}{afb}{FB}
\FXRegisterAuthor{sm}{asm}{SM}
\FXRegisterAuthor{hu}{ahu}{HU}

\title{On Language Varieties Without Boolean Operations}

\author{%
  Fabian Birkmann
  \and
  Stefan Milius
  \and
  Henning Urbat
}
\authorrunning{F.~Birkmann, S.~Milius, H.~Urbat}
\institute{Friedrich-Alexander-Universität
  Erlangen-Nürnberg, Germany\\
  \email{\{fabian.birkmann,stefan.milius,henning.urbat\}@fau.de}%
}

\maketitle

\begin{abstract}
  Eilenberg's variety theorem marked a milestone in the algebraic theory of regular languages by establishing a formal correspondence between properties of regular languages and properties of finite monoids recognizing them. Motivated by classes of languages accepted by quantum finite automata, we introduce \emph{basic varieties of regular languages}, a weakening of Eilenberg's original concept that does not require closure under any boolean operations, and prove a variety theorem for them. To do so, we investigate the algebraic recognition of languages by \emph{lattice bimodules}, generalizing Kl\'ima and Pol\'ak's lattice algebras, and we utilize the duality between algebraic completely distributive lattices and posets.
\end{abstract}


\section{Introduction}
The introduction of algebraic methods into the study of regular
languages pro\-vides a convenient classification system that allows to
study finite automata and their languages in terms of associated
finite algebraic structures. A celebrated example is Schützenberger's
theorem~\cite{mps} stating that a language is star-free iff its
syntactic monoid is aperiodic, thus proving the decidability of
star-freeness. Eilenberg's \emph{variety theorem}~\cite{eil}
formalizes this type of correspondence as a bijection between
\emph{varieties of regular languages} (i.e.\ classes of regular
languages closed under the set-theoretic boolean operations, word
derivatives and preimages of monoid homomorphisms) and
\emph{pseudovarieties of monoids} (i.e.\ classes of finite monoids
closed under finite products, submonoids and quotient
monoids).

Numerous extensions and generalizations of Eilenberg's theorem have
been discovered over the past four decades, differing from the
original one by either changing the type of languages under
consideration, e.g.\ from regular languages to $\omega$-regular
languages~\cite{wilke91}, or by considering notions of varieties with
relaxed closure properties. On the algebraic side, such a relaxation
requires to replace monoids by more complex algebraic structures. For
instance, Pin~\cite{pin-1995} studied \emph{positive varieties of
  regular languages}, where the closure under complement is dropped,
and proved them to biject with pseudovarieties of \emph{ordered}
monoids. Subsequently, Pol\'ak~\cite{polak2001} introduced
\emph{disjunctive varieties of regular languages}, where in addition
to closure under complement also the closure under intersection is
dropped, and related them to pseudovarieties of idempotent semirings.

One item is conspicuously missing from this list: a variety theorem
for classes of languages that need not be closed under any boolean
operations, i.e.\ in which only closure under word derivatives and
preimages of monoid homomorphisms is required. Such \emph{basic
  varieties of regular languages} subsume all the above notions of
varieties and naturally arise in several areas of automata theory,
most notably in the study of languages accepted by reversible finite
automata~\cite{gol-pin} or quantum finite automata~\cite{kon-wat}. In
the present paper, we close this gap by developing the theory of basic
varieties. As the corresponding algebraic structure we introduce
\emph{lattice bimodules}, a two-sorted generalization of the
\emph{lattice algebras} recently studied by Kl\'ima and
Pol\'ak~\cite{kli-pol}, as algebraic recognizers for regular
languages. The two-sorted approach allows for a clearer and more
conceptual view of the underlying categorical and universal algebraic
concepts. As our main result, we establish the following algebraic
classification of basic varieties:

\begin{bvthm}
  Basic varieties of regular languages correspond bijectively to pseudovarieties
  of lattice bimodules.
\end{bvthm}

\noindent This answers the open problem of Kl\'ima and Pol\'ak~\cite{kli-pol} about
an Eilenberg-type correspondence.
Our presentation of the theorem and
its proof is inspired by the recently developed duality-theoretic
perspective on algebraic language theory~\cite{ggp08,Salamanca16,uacm17,ammu19}, which provides the
insight that correspondences between language varieties and
pseu\-do\-va\-rieties of algebraic structures can be understood in
terms of an underlying dual equivalence of categories. In our setting,
we shall demonstrate that pseudovarieties of lattice bimodules can be
interpreted as \emph{theories} of lattice bimodules in the category
$\AlgCDL$ of algebraic completely distributive lattices, while basic
varieties give rise to (basic) \emph{cotheories} of regular languages in the
category $\POS$ of posets. Our Eilenberg correspondence for basic
varieties then boils down to an application of the well-known dual
equivalence
\[
  \AlgCDL \simeq^\op \POS.
\]
Let us note that our main result is not an instance of previous
category-theoretic generalizations of Eilenberg's
theorem~\cite{Salamanca16,uacm17,ammu19,boj} since the two-sorted
nature of lattice bimodules requires to introduce the novel concept of
\emph{reduced} structures, which makes the ensuing notion of
pseudovariety more intricate than the ones studied in
\emph{op.\,cit}. However, much of the general methodology developed
there turns out to apply smoothly, which can be seen as further
evidence of its scope and flexibility. In order to make the present
paper accessible to readers not familiar with the previous work, we
opted to give a self-contained presentation of our results, merely
assuming some familiarity with basic category theory.

\section{Lattice Bimodules} \label{sec:lattice-bimodules}
In this section we introduce a new algebraic structure whose aim it is
to capture languages varieties that are not necessarily closed under boolean operations. Our notion is a two-sorted generalization of Kl\'ima and Pol\'ak's \emph{lattice algebras}~\cite{kli-pol}. Intuitively, for our intended
purpose the following structure should be present:
%
\begin{enumerate}
\item
  a \emph{monoid action} that corresponds to word derivation on the language side;
\item
  \emph{lattice-like operations} to compensate for the missing closure under union and intersection on the language side;
\item
  equational axioms specifying the interaction of (1) and (2). 
  
\end{enumerate}

\noindent
From a categorical perspective, the last point means that our algebras can be modeled by a \emph{monad}. This allows us to use previous work on languages recognizable by monad algebras \cite{boj,Salamanca16,uacm17} as a guide towards our results.




While Klíma and Polák considered distributive lattices with an
embedded monoid acting on them, we upgrade the lattice to a
\emph{completely distributive lattice} (shortly, \emph{CDL}), i.e.\ a
complete lattice satisfying the infinite distributive law
$\bigvee_{i\in I} \bigwedge_{j\in J_i} x_{i,j} = \bigwedge_{f\in F}
\bigvee_{i\in I} x_{i,f(i)}$ for every family
$\{ x_{i,j} : i\in I,\, j\in J_i \}$ of elements, where $F$ is the set
of all choice functions $f$ mapping each $i\in I$ to some
$f(i)\in J_i$. \emph{Morphisms of CDLs} are maps preserving all joins
and meets. We let $\CDL$ denote the category of CDLs and their
morphisms. Even though completeness makes no difference for finite
structures, completely distributive lattices admit a more convenient
duality theory than general distributive lattices.

In addition, in lieu of an embedded monoid we use a two-sorted
structure with a monoid in the first sort. This avoids partial
operations, which are somewhat awkward from the perspective of
(categorical) universal algebra.

\begin{defn}
\begin{enumerate}
\item A \emph{lattice bimodule} $(M,D,\iota,\tr,\tl)$, abbreviated as
  $(M,D)$, is given by a monoid $(M, ·, 1)$, a CDL $(D, ∨, ∧)$, and
  three operations
  \[
    \tr\: M \times D \rightarrow D,\qquad
    \tl \: D \times M \rightarrow D, \qquad \iota\colon M\to D,
  \]
  such that $\tr$ and $\tl$ form a monoid biaction of $M$ on $D$ that
  distributes over the lattice operations, and $\iota$ translates the
  multiplication of $M$ to $\tl$ and $\tr$; that is, for all
  $m, n ∈ M$, $d \in D$ and $\{d_i\}_{i \in I} \subseteq D$, the
  following equational laws hold:
  \[
    \begin{array}{rcl@{\qquad}rcl}
        (m · n) \tr d &  = & m \tr (n \tr d), &
        d \tl (m · n) & = & (d \tl m) \tl n,
      \\
	  1 \tr d & = & d,  &  
	  d \tl 1 & = & d, 
	\\
	(m \tr d) \tl n & = & m \tr (d \tl n),
	\\
      m \tr (⋁_{i ∈ I} d_i) & = & ⋁_{i ∈ I} (m \tr d_i), & 
      (⋁_{i ∈ I} d_i) \tl m & = & ⋁_{i ∈ I} (d_i \tl m),\\
      m \tr (⋀_{i ∈ I} d_i) & = & ⋀_{i ∈ I} (m \tr d_i), &
      (⋀_{i ∈ I} d_i) \tl m & = & ⋀_{i ∈ I} (d_i \tl m),
      \\
	  m \tr ι(n) & = & ι(m · n), &
      ι(m) \tl n & = & ι(m · n).
      \\	  
    \end{array}
  \]
\takeout{
\begin{itemize}
        \item $1 \tr d = d = d \tl 1$
        \item $(m · n) \tr d = m \tr (n \tr d)$,
            $d \tl (m · n) = (d \tl m) \tl n$
        \item $m \tr ι(n) = ι(m · n) = ι(m) \tl n$
        \item $m \tr (d \tl n) = (m \tr d) \tl n $
        \item $m \tr (⋀_{i ∈ I} d_i) = ⋀_{i ∈ I} (m \tr d_i)$,
              $(⋀_{i ∈ I} d_i) \tl m = ⋀_{i ∈ I} (d_i \tl m)$
        \item $m \tr (⋁_{i ∈ I} d_i) = ⋁_{i ∈ I} (m \tr d_i)$,
              $(⋁_{i ∈ I} d_i) \tl m = ⋁_{i ∈ I} (d_i \tl m)$
    \end{itemize}
  } Note that since the least and the greatest element of $D$ are
  given by $\bot=\bigvee \emptyset$ and $\top=\bigwedge \emptyset$,
  resp., we also have $m \tr ⊥ = ⊥ = ⊥ \tl m$ and
  $m \tr ⊤ = ⊤ = ⊤ \tl m$.
  
\item  A \emph{homomorphism} from a lattice bimodule $(M, D, ι, \tr,
  \tl)$ to a lattice bimodule $(M', D', ι', \tr', \tl')$ is given by a two-sorted map
  $h=(h^\star,h^\diamond) \: (M,D) \to(M',D')$ such that $h^\star$ is a monoid
  homomorphism, $h^\diamond$ is a morphism of completely distributive
  lattices and the following diagrams commute:
 \begin{equation*}
 \begin{tikzcd}[row sep=1.6em] 
   M\times D
   \arrow[r, "\tr"]
   \arrow[d, "h^\star\times h^\diamond"']
   &
   D
   \arrow[d, "h^\diamond"]
   \\
   M'\times D'
   \arrow[r, "\tr'"]
   &
   D'
 \end{tikzcd}\qquad
 \begin{tikzcd}[row sep=1.6em]
   D\times M
   \arrow[r, "\tl"]
   \arrow[d, "h^\diamond \times  h^\star"']
   &
   D
   \arrow[d, "h^\diamond"]
   \\
   D' \times M'
   \arrow[r, "\tl'"]
   &
   D'
 \end{tikzcd}
\qquad
 \begin{tikzcd}[row sep=1.6em]
   M
   \arrow[r, "\iota"]
   \arrow[d, "h^\star"']
   &
   D
   \arrow[d, "h^\diamond"]
   \\
   M'
   \arrow[r, "\iota'"]
   &
   D'
 \end{tikzcd}
 \end{equation*}
\emph{Subbimodules} and \emph{quotient bimodules} of lattice bimodules are represented by sortwise injective and surjective homomorphisms, respectively.
\end{enumerate}
\end{defn}
\takeout{
\begin{rem}\label{rem:latticalg}
Lattice bimodules as closely related to Kl\'ima and Pol\'ak's \emph{lattice algebras}~\cite{kli-pol}. In the definition of the latter, the lattice $D$ is not necessarily complete and $\tr$ and $\tl$  distribute only over finite, not arbitrary, joins and meets. In addition, the monoid $M$ is required to be a subset of $D$ and the map $\iota\: M\monoto D$ is the inclusion. In our setting of lattice bimodules, this condition will naturally arise from categorical considerations when we study \emph{reduced} lattice bimodules, see \autoref{subsec:generated-injective-reduced}.
\end{rem}}
We let $\LBM$ denote the category of lattice bimodules and their homomorphisms.

A \emph{free} lattice bimodule over a pair $({\Sigma}, {\Gamma})$ of
sets is given by a lattice bimodule $(\hat \Sigma, \hat \Gamma)$
together with a sorted map
$η=(\eta^\star,\eta^\diamond)\:
(\Sigma,\Gamma)→(\hat{\Sigma},\hat{\Gamma})$ satisfying the universal
mapping property: for every sorted map
$h_0 \: (\Sigma, \Gamma) → (M, D)$ to a lattice bimodule $(M, D)$
there exists a unique lattice bimodule homomorphism
$h\: (\hat{\Sigma}, \hat{\Gamma}) \rightarrow (M, D)$ such that
$h\cdot \eta = h_0$.
In the following, we denote by $\fm \Sigma$ the free monoid on the set
$\Sigma$ with neutral element $\epsilon \in \fm \Sigma$ and by
$\fcdl(\Gamma)$ the free completely distributive lattice~\cite{mar} on
the set $\Gamma$. The latter can be described as the lattice of
downwards closed subsets of the power set $\Pow(\Gamma)$, or
equivalently as the lattice of all formal expressions
$\bigvee_{i\in I} \bigwedge_{j\in J_i} x_{i,j}$, where
$x_{i,j}\in \Gamma$, modulo the equational laws of CDLs. We view
$\Gamma$ as a subset of $\fcdl(\Gamma)$.
\begin{proposition}\label{prop:lbm-free}
  The free lattice bimodule over $(\Sigma, \Gamma)$ is given by
  $\eta\: (\Sigma,\Gamma) \to (\fm{\Sigma}, \flb{\Sigma}{\Gamma})$
  with $\eta^\star(a)=a$,
  $\eta^\diamond(b) = (\epsilon, b, \epsilon)$, and operations
  uniquely determined by the following identities for
  $u,v,w\in \Sigma^\star$ and $z\in \Gamma$:
\[ \iota(u)=u,\; u\tr v = uv,\; u\tr (v,z,w)=(uv,z,w),\; u\tl v = uv,\; (v,z,w)\tl u = (v,z,wu). \]
\end{proposition}
\begin{notation}
  We write $\f[\Sigma]=(\Sigma^\star,\fcdl(\Sigma^\star))$ for the
  free lattice bimodule on $(\Sigma, \emptyset)$. Note that a
  homomorphism $h\colon \f[\Sigma]\to (M,D)$ is completely determined
  by its first component
  $h^\star\colon \Sigma^\star\to M$. In fact, its second component
  $h^\diamond\colon \Sigma^\diamond \to D$ is the unique
  $\CDL$-morphism extending the map $\iota\cdot h^\star\colon \Sigma^*\to D$.
\end{notation}
\takeout{
\subsection{⭑-generated, ⭑-embedded and reduced  lattice bimodules}\label{subsec:generated-injective-reduced}
We now examine three properties of lattice bimodules, namely
\emph{⭑-embeddedness}, \emph{⭑-generatedness} and
\emph{reducedness}. The latter will play a key role in definition of a
pseudovariety in \autoref{sec:pseudovarieties}.}
We now define three properties of lattice bimodules needed subsequently. The last one will play a key role in the study of pseudovarieties in \autoref{sec:pseudovarieties}.
\newcommand{\distrib}[1]{⋁_{j \in J}⋀_{k \in K_j}{#1} }
\begin{defn}\label{def:gen_emb_red} A lattice bimodule $(M, D)$ is called
  \begin{enumerate}
  \item \emph{⭑-generated} if the
    complete lattice $D$ is generated by the image $\iota[M]\seq D$:
     For all $d ∈ D$ there exist elements $m_{i,j} ∈ M$ such that
    $d=\bigvee_{i\in I} \bigwedge_{j\in J_i} \iota(m_{i,j})$;
  \item \emph{⭑-embedded} if the operation $ι\: M → D$ is injective;
    
  \item \emph{reduced} if for every
    quotient bimodule $h \: (M, D) ↠ (M', D')$ such that $h^\diamond\colon D\epito D'$ is a $\CDL$-isomorphism, $h$ is an $\LBM$-isomorphism.
  \end{enumerate}
\end{defn}
Intuitively, reducedness captures lattice bimodules that carry information only in
their second component by demanding them to be ``as minimal as possible'' in
their first component,  here expressed through a characterization of
quotients. Finite $\star$-embedded lattice bimodules are precisely the finite lattice algebras of Kl\'ima and Pol\'ak~\cite{kli-pol}. The following lemma links the above concepts:

\takeout{
The operation $\iota\: M \to D$ of a lattice bimodule should be
understood as an ``embedding'' of the monoid $M$ in the lattice $D$.
We are interested in the case where the lattice is ``generated'' by
the monoid, in the sense that any homomorphism with codomain $(M, D)$
is determined uniquely by its ⭑-component.}


\begin{lemma}\label{lem:reduced-vs-embedded}
\begin{enumerate}
\item A lattice bimodule $(M,D)$ is $\star$-generated if and only if there exists a surjective homomorphism from  $\f[\Sigma]$ to $(M,D)$ for some set $\Sigma$. 
\item Every $\star$-embedded lattice bimodule is reduced.
\item Every $\star$-generated reduced lattice bimodule is $\star$-embedded.
\end{enumerate}
\end{lemma}
\newcommand{\e}[1]{\equiv_{#1}} In the categorical approach to variety
theorems~\cite{uacm17} it was shown that the key to
understanding language derivatives lies in the concept of a
\emph{unary presentation} of an algebraic structure. Informally, such
a presentation expresses the structure of an algebra in terms of
suitable unary operations in the underlying category, which then
dualize to the derivative operations on the set of languages
recognized by that algebra. The heterogeneous nature of our present setting, which regards lattice bimodules as algebraic structures over the product category $\Set\times \CDL$, requires a slight adaptation of the concepts from \emph{op.\ cit}.
\begin{defn}\label{D:up}
  Let $(M,D)$ be a lattice bimodule. A \emph{unary operation} on
  $(M,D)$ is either a map of type $M \to M$ or $M \to D$, or a $\CDL$-morphism $D \to D$. A set $\U$ of unary operations forms a \emph{unary
    presentation} of $(M,D)$ if for every pair
  $e=(e^\star, e^\diamond)$ of a surjective map
  $e^\star\colon M\epito {M'}$ and a surjective $\CDL$-morphism
  $e^\diamond \colon D\epito {D'}$, the following statements are
  equivalent:
\begin{enumerate}
\item The map $e$ carries a quotient bimodule of $(M,D)$, i.e.~there exists a lattice bimodule structure on $({M'},{D'})$ making $e$ a homomorphism of lattice bimodules.
\item For every $u\in \U$, there exists $\bar{u}$ making the respective square below commute:
\begin{equation*}
  \begin{tikzcd}
    M
    \arrow[d, "e^\star"', two heads]
    \arrow[r, "u"]
    &
    M
    \arrow[d, "e^\star", two heads]
    \\
    {M'}
    \arrow[r, "\bar{u}", dashed]
    &
    {M'}
  \end{tikzcd}
\qquad 
  \begin{tikzcd}
    M
    \arrow[d, "e^\star"', two heads]
    \arrow[r, "u"]
    &
    D
    \arrow[d, "e^\diamond", two heads]
    \\
    {M'}
    \arrow[r, "\bar{u}", dashed]
    &
    {D'}
  \end{tikzcd}
\qquad 
  \begin{tikzcd}
    D
    \arrow[d, "e^\diamond"', two heads]
    \arrow[r, "u"]
    &
    D
    \arrow[d, "e^\diamond", two heads]
    \\
    {D'}
    \arrow[r, "\bar{u}", dashed]
    &
    {D'}
  \end{tikzcd}
\end{equation*}
\end{enumerate} 
\end{defn}

%
%
%

\begin{lemma}\label{L:up}
  Every lattice bimodule $(M, D)$ admits a unary presenation composed of the following unary operations ranging over $m \in M$ and $d \in D$:
  \[
    (m\, ·) , (·\, m) \: M \to M,\hspace{6pt} ι \: M \to D,\hspace{6pt}
    (m \,\triangleright) , (\triangleleft \, m) \: D \to D,\hspace{6pt}
    (\triangleright \, d), (d \, \triangleleft) \: M \to D.
  \]
\end{lemma}
\noindent
Note that the maps $(m \,\triangleright)$ and $(\triangleleft \, m)$ are indeed $\CDL$-morphisms, as required.

\section{Pseudovarieties of Reduced Lattice Bimodules}\label{sec:pseudovarieties}


In this section, we introduce \emph{pseudovarieties} and \emph{theories} of (reduced) lattice bimodules and show them to be in one-to-one correspondence. The concept of a pseudovariety originates in Eilenberg's classical variety theorem~\cite{eil} where a \emph{pseudovariety of monoids} is a class of finite monoids closed under finite products, submonoids, and quotient monoids. In our setting of lattice bimodules, we shall consider pseudovarieties of $\star$-generated reduced lattice bimodules. Their definition is slightly more involved than in the case of monoids because subbimodules of $\star$-generated lattice bimodules are not necessarily $\star$-generated and quotient bimodules of reduced lattice bimodules are not necessarily reduced.

\begin{defn}\label{D:pseudovar}
  A \emph{pseudovariety of lattice bimodules} is a class $\V$ of $\star$-generated reduced finite lattice bimodules such that
  \begin{enumerate}
  \item\label{D:pseudovar:2}
    $\V$ is closed under reduced quotients: for every surjective homomorphism $e\colon (M,D)\epito (M',D')$ of lattice bimodules, if $(M, D) ∈ \V$ and $(M',D')$ is reduced then $(M',D')\in \V$.
  \item\label{D:pseudovar:3}
    $\V$ is closed under $\star$-generated subbimodules of finite products: for every injective homomorphism $(M, D) \rightarrowtail ∏_{i=1}^n (M_i, D_i)$ of lattice bimodules, if $(M_i,D_i)\in \V$ for $i=1,\ldots,n$ and $(M,D)$ is $\star$-generated then $(M,D)\in \V$.
\end{enumerate}
\end{defn}
We shall also consider the related notion of a \emph{local}
pseudovariety. It is local in the sense that it involves only quotient
bimodules of a fixed free lattice bimodule $\f[\Sigma]$. The set of
all such quotients carries a natural partial order: given
$e_i\colon \f[\Sigma]\epito (M_i,D_i)$, $i=0,1$, we put $e_0\leq e_1$
iff $e_0=h\cdot e_1$ for some $h$.
\begin{defn}
  A \emph{local pseudovariety of lattice bimodules} over the finite
  set $\Sigma$ is a set $\locT$ of quotient bimodules of
  $\f[\Sigma]$ such that
  \begin{enumerate}
  \item The codomain of every $e \in \locT$ is finite and reduced. (Note that it is also $\star$-generated by \autoref{lem:reduced-vs-embedded}(1).)

  \item $\locT$ is downwards closed: if $e ∈ \locT$ and
    $e' \: \f ↠ (M, D)$ is a quotient bimodule with reduced codomain, then $e'\leq e$ implies
    $e' ∈ \locT$.

  \item $\locT$ is directed: if $e_0, e_1 ∈ \locT$, then there
    exists $e ∈ \locT$ with $e_0, e_1 ≤ e$.
  \end{enumerate}
\end{defn}
\noindent
In order-theoretic terminology, a local pseudovariety is thus
precisely an ideal in the poset of finite reduced quotient bimodules
of $\f[\Sigma]$.

\enlargethispage{11pt}
\begin{defn}\label{D:theory}
  A \emph{theory of lattice bimodules} is a family
  $\Th = (\locT)_{\Sigma\in \A}$ of local pseudovarieties, 
with $\Sigma$ ranging over the class $\A$ of finite sets, such that for each
  homomorphism $h \: \f[Δ] → \f[Σ]$ and $e_\Sigma ∈ \locT[Σ]$ their composite
  $e_\Sigma·h$ \emph{lifts through \locT[Δ]}, that is, there exist
  $e_\Delta \in \locT[\Delta]$ and $\overline{h}$ such that $e_\Sigma
  \cdot h = \bar h \cdot e_\Delta$.
\begin{equation*}
  \begin{tikzcd}[row sep=1.8em]
    \f[Δ]
    \arrow[d, "e_\Delta"', two heads, dashed]
    \arrow[r, "h"]
    &
    \f[Σ]
    \arrow[d, "e_\Sigma", two heads]
    \\
    {(M', D')}
    \arrow[r, "\bar{h}", dashed]
    &
    {(M, D)}
  \end{tikzcd}
\end{equation*}
\end{defn}
\begin{notation}\label{not:vt-tv}
  \begin{enumerate}
  \item Given a theory $\Th$, let
    $\mcpower{V}{T}$ be the class of all lattice bimodules $(M,D)$ such that some $\locT[\Sigma]$ contains a quotient with codomain $(M,D)$.
  \item Given a pseudovariety $\V$, form the family
    $\mcpower{T}{V} = (\mcpower{T}{V}_Σ)_{\Sigma\in \A}$ where
    $\mcpower{T}{V}_Σ$ consists of all quotient bimodules of $\f[\Sigma]$ with codomain in $\V$.
  \end{enumerate}
\end{notation}
The class of all pseudovarieties of lattice bimodules forms a
lattice ordered by inclusion. Similarly, the class of all theories of
lattice bimodules forms a lattice ordered by pointwise inclusion:
$\Th \leq \Th'$ iff
$\Th_\Sigma\seq \Th_\Sigma'$ for each
$\Sigma$.

\begin{theorem}\label{thm:iso-lbm-rt}
  The maps $\V ↦ \mcpower{T}{V}$ and
  $\Th ↦ \mcpower{V}{T}$ give rise to an isomorphism between the lattice of
  pseudovarieties of lattice bimodules and 
  the lattice of theories of lattice bimodules.
\end{theorem}
We conclude this section with another characterization of theories,
linking them to the concept of a unary presentation. For any set Σ,
let $\mathbb{U}_Σ$ be the canonical unary presentation of the free
lattice bimodule \f given by \autoref{L:up}, and denote by
$\overline{\mathbb{U}}_Σ$ its closure under
composition. Then \barU also forms a unary presentation of \f. We
write $\barU(S, T) ⊆ \barU$ for the set of unary operations in \barU
with domain $S$ and codomain $T$, where
$S,T\in \{ \Sigma^\star,\Sigma^\diamond\}$.  In particular,
$\barU(\Sigma^⋄,\Sigma^⋄) = \{x ↦ vxw \pipe v, w ∈ Σ^⭑ \}$.
\begin{defn}\label{def:uquot}
\begin{enumerate}
\item A quotient $e \: Σ^⋄ ↠ D$ in \CDL is called a \emph{\U-quotient}
  if for every unary operation $u ∈ \barU(\Sigma^⋄,\Sigma^⋄)$ there
  exists a $\CDL$-morphism $\bar{u}\colon D\to D$ such that
  $e·u=\bar{u}·e$.  \takeout{
    \begin{equation*}
      \begin{tikzcd}
        Σ^⋄
        \arrow[r, "u"]
        \arrow[d, "e", two heads]
        &
        Σ^⋄
        \arrow[d, "e", two heads]
        \\
        D
        \arrow[r, "\bar{u}"]
        &
        D
      \end{tikzcd}
    \end{equation*}}
  We call such a $\bar{u}$ a \emph{lifting of $u$ along $e$}.
  
\item A \emph{local pseudovariety of $\U$-quotients} over the finite
  set $\Sigma$ is an ideal in the poset of finite $\U$-quotients of
  $\Sigma^\diamond$.
  
\item A \emph{theory} of \U-quotients is a family $\Th = \T$ of local
  pseudovarieties of $\U$-quotients such that for each lattice
  bimodule homomorphism $h \: \f[Δ] → \f[Σ]$ and $e_\Sigma ∈ \locT[Σ]$
  their composite $e_\Sigma·h^⋄$ \emph{lifts through \locT[Δ]}: there
  exist morphisms $e_\Delta \in \locT[\Delta]$ and $\overline{h}$ such
  that $e_\Sigma \cdot h^\diamond = \overline{h}\cdot e_\Delta$.
  \begin{equation}\label{eq:utheory}
  \begin{tikzcd}
    \Sigma^⋄
    \arrow[d, "e"', two heads]
    \arrow[r, "u"]
    &
    Σ^⋄
    \arrow[d, "e", two heads]
    \\
    D
    \arrow[r,  "\overline{u}", dashed]
    &
    D
  \end{tikzcd}
  \qquad\qquad
  \begin{tikzcd}[row sep=1.6em]
    Δ^⋄
    \arrow[d, "e_\Delta"', two heads]
    \arrow[r, "h^⋄"]
    &
    Σ^⋄
    \arrow[d, "e_\Sigma", two heads]
    \\
    D'
    \arrow[r, "\bar{h}", dashed]
    &
    D
  \end{tikzcd}
\end{equation}
\end{enumerate}
\end{defn}
 
\begin{proposition}\label{prop:iso-thUquot-thLBM}
  The lattice of theories of lattice bimodules is isomorphic to the lattice of theories of \U-quotients. The isomorphism is given by $\Th \mapsto \Th^\diamond$, where $\Th^\diamond$ consists of all quotients in $\Th$ restricted to their $\diamond$-component.
\end{proposition}
 The advantage in using theories of \U-quotients is that they are easier to dualize but still carry as much information as theories of lattice bimodules.

\section{Basic Varieties of Regular Languages}\label{sec:languages}

In this section, we study lattice bimodules as recognizers for regular
languages. Their purpose is to capture classes of regular languages with no
boolean closure at all, which we thus call \emph{basic
  varieties}. Observe that since the set $2=\{0,1\}$ with $0\leq 1$
forms a CDL and the set $\Sigma^\star$ generates the free completely
distributive lattice $Σ^⋄$, we get the correspondence
$\Pow(Σ^⭑) \cong \Set(Σ^⭑, 2) \cong \CDL(Σ^⋄, 2)$. We use the term
``language'' for elements of any of these sets, identifying elements
that correspond to each other via the bijections. Thus, we use the
same symbol for a subset $L ⊆ Σ^⭑$ and for its characteristic
function. We denote the extension of $L \: Σ^\star → 2$ to a lattice
morphism by $L^⋄ \: Σ^⋄ → 2$, and in turn denote the restriction of a
lattice morphism $L \: Σ^⋄ → 2$ to $Σ^⭑$ by
$L^⭑ = L · ι \: Σ^⭑ → Σ^⋄ → 2$.

\newcommand{\hatL}{\hat{L}}
\begin{defn}\label{def:lang-rec} 
A language $L \: Σ^⭑ → 2$ is \emph{recognized} by a finite lattice
  bimodule $(M, D)$ if there exists a lattice bimodule homomorphism
  $h \: \f → (M,D) $ and a $\CDL$-morphism $p \: D → 2$ with
  $L^⋄ = p · h^⋄ $. 
  \[
   \begin{tikzcd}[row sep=1.4em]
     Σ^⋄
     \arrow[d, "h^⋄"']
     \arrow[rd, "L^⋄", bend left=20]
     &
     \\
     D
     \arrow[r, "p"]
     &
     2
   \end{tikzcd}
 \]
\end{defn} 

\begin{lemma}\label{lem:recognizes-regular}
    The languages recognizable by finite lattice bimodules are precisely the regular languages.
\end{lemma}
Recall that pseudovarieties of lattice bimodules consist of
$\star$-generated reduced bimodules. This restriction does not limit the recognized languages:
\begin{lemma}\label{cor:reduced-generated-recognizes-regular}
  Every language $L$ recognizable by a finite lattice bimodule is
  recognizable by a finite ⭑-generated reduced lattice bimodule.
\end{lemma}
%
We now introduce our concept of a language variety that we will show
to correspond to pseudovarieties of lattice bimodules. It
subsumes Eilenberg's original concept~\cite{eil}, as well as its variants due to Pin~\cite{pin-1995} and Pol\'ak~\cite{polak2001}, by dropping the
requirement of being closed under any set-theoretic boolean
operations. Recall that the \emph{derivatives} of a language $L \subseteq \Sigma^\star$ are the languages
 $v^{-1}Lw^{-1} = \{u\in \Sigma^\star \mid vuw \in L\}$ for $v, w \in \Sigma^\star$. The \emph{preimage} of $L$ w.r.t.\ a
monoid homomorphism $g\: \Delta^\star \to \Sigma^\star$ is given by $g^{-1}L = \{w \in \Delta^\star \mid g(w) \in L\}$.
In the following we write \reg{Σ} for the set of all regular languages
over $\Sigma$.

\begin{defn}\label{language-variety}
  \begin{enumerate}
  \item
    A \emph{local basic variety of languages over $Σ$} is a
    set $V_Σ ⊆ \reg{Σ}$ closed under derivatives: If $L ∈ V_Σ$ then
    $v^{-1}Lw^{-1} ∈ V_Σ$ for all $v,w\in\Sigma^\star$.
  \item A \emph{basic variety of languages} is a family
    $(V_Σ\,\seq\, \reg{\Sigma})_{Σ ∈ \A}$ of local varieties closed
    under preimages of monoid homomorphisms: If $L ∈ V_Σ$ then
    $g^{-1}L ∈ V_Δ$ for each monoid homomorphism
    $g\colon \Delta^\star\to \Sigma^\star$.
  \end{enumerate}
\end{defn}
Just as pseudovarieties of reduced lattice bimodules can be presented
as theories, basic varieties of languages correspond uniquely to
\emph{cotheories}. In the following definition, $\Pow(X)$ denotes the poset of subsets of a set $X$. Recall that an ideal of $\Pow(X)$ is a subset $I\seq\Pow(X)$ that is downwards closed and upwards directed. 
\begin{defn}\label{def:cotheory}
  A \emph{basic cotheory of regular languages} is a family
\[T = (I_\Sigma \seq \Pow(\reg{\Sigma}))_{Σ ∈ \A}\]
of ideals with the following properties:
\begin{enumerate}
\item\label{def:cotheory:1} Every element $F_\Sigma\in I_\Sigma$ is a finite basic local variety.
\item\label{def:cotheory:2} $T$ is closed under preimages of monoid homomorphisms: If
  $F_\Sigma ∈ I_Σ$, then $g^{-1}[F_\Sigma] = \{  g^{-1}L\pipe L\in
  F_\Sigma\}\in I_\Delta$ for each monoid homomorphism $g\: \Delta^\star\to \Sigma^\star$.
\end{enumerate}
\end{defn}
In diagrammatic terms, \ref{def:cotheory:1} means that for every
$u\in \U_\Sigma(\Sigma^\diamond,\Sigma^\diamond)$, viewed as a map
$u\colon \Sigma^\star\to \Sigma^\star$ by restricting its domain and
codomain, the preimage map
$u^{-1}\colon \Pow(\Sigma^\star)\to \Pow(\Sigma^\star)$ restricts to
$F_\Sigma$. Indeed, since $\U_\Sigma(\Sigma^\diamond,\Sigma^\diamond)$
consists of all unary operations $u$ of the form $x\mapsto vxw$ for
$v,w\in \Sigma^\star$, the map $u^{-1}$ is given by
$L\mapsto v^{-1}Lw^{-1}$. Similarly, \ref{def:cotheory:2} means that for every
$F_\Sigma\in I_\Sigma$ and $g\: \Delta^\star\to \Sigma^\star$, the map
$g^{-1}\colon \Pow(\Sigma^\star)\to \Pow(\Delta^\star)$ restricts to
one between $F_\Sigma$ and some $F_\Delta\in I_\Delta$.
  \begin{equation}\label{eq:cotheory}
    \begin{tikzcd}
      \Pow(Σ^⭑)
      \arrow[r, "u^{-1}"]
      &
      \Pow(\Sigma^⭑)
      \\
      F_Σ
      \arrow[u, "\seq", hook]
      \arrow[r, dashed]
      &
      F_\Sigma
      \arrow[u, "\seq"', hook]
    \end{tikzcd}
\qquad\qquad
    \begin{tikzcd}
      \Pow(Σ^⭑)
      \arrow[r, "g^{-1}"]
      &
      \Pow(Δ^⭑)
      \\
      F_Σ 
     \arrow[u, "\seq", hook]
      \arrow[r, dashed]
     &
      F_Δ
      \arrow[u, "\seq"', hook]
    \end{tikzcd}
  \end{equation}
\noindent
Basic varieties of regular languages form a lattice ordered by inclusion. Similarly,
 basic cotheories of regular languages are ordered by pointwise inclusion.

\begin{theorem}\label{thm:iso-thlang-varlang}
  The lattice of basic varieties of regular languages is isomorphic to
  the lattice of basic cotheories of regular languages. The isomorphism and its inverse are
  given pointwise for $Σ ∈ \A$ by the maps
  \begin{align*}
    V_Σ &↦ \{\, F \seq V_\Sigma  \mid F \textrm{ is a finite basic local subvariety of $V_Σ$}\,\}\;\;\;\text{and}\;\;\;     I_Σ ↦ \bigcup I_Σ.\\
  \end{align*}
\end{theorem}

\section{Duality and the Basic Variety Theorem}\label{sec:variety-theorem}
  \tikzset{
    subs/.style={anchor=south, rotate=90, inner sep=.5mm}
  }

\newcommand{\homcdl}[0]{\ensuremath{\CDL(-, 2)}\xspace}
\newcommand{\hompos}[0]{\ensuremath{\POS(-, 2)}\xspace}

\newcommand{\ds}[1]{\ensuremath{\downarrow\!#1}\xspace}
\newcommand{\us}[1]{\ensuremath{\uparrow\!#1}\xspace}


The glue between the algebraic concepts of
\autoref{sec:pseudovarieties} and the language-theoretic ones of
\autoref{sec:languages} is provided by \emph{duality}, more precisely,
the dual equivalence $\AlgCDL \simeq^\op \POS$ between the full
subcategory $\AlgCDL$ of $\CDL$ given by \emph{algebraic} completely
distributive lattices and the category $\POS$ of posets and monotone
maps~\cite{dav-pri}. Observe that since all free CDLs and finite CDLs
are algebraic, a theory of $\U$-quotients (\autoref{def:uquot}) lives
in the category $\AlgCDL$. Similarly, a basic cotheory of regular
languages (\autoref{def:cotheory}) lives in $\POS$, viewing the set
$\Pow(\Sigma^\star)$ of languages as a poset ordered by inclusion. Let
us now make the key observation that, under the above duality,
theories of $\U$-quotients \emph{dualize} to basic cotheories of
regular languages: One can show that, up to isomorphism, the duals of
the commutative squares \eqref{eq:utheory} in $\AlgCDL$ are precisely
the commutative squares \eqref{eq:cotheory} in $\POS$ where
$g=h^\star$ and $F_\Sigma$ and $F_\Delta$ are the posets of languages
recognized by $e_\Sigma$ and $e_\Delta$, respectively. We can
therefore bring the results of the previous sections together to
establish our main result: 
\begin{theorem}[Basic Variety Theorem]\label{variety-theorem}
  The lattice of basic varieties of regular languages is isomorphic to the lattice of pseudovarieties of lattice bimodules.
\end{theorem}
\proof
We simply compose all the previously established lattice isomorphisms:
\begin{align*}
  \phantom{\cong}
  &\textrm{ Pseudovarieties of lattice bimodules}
  \\
  \cong
  &\textrm{ Theories of lattice bimodules}
  & \text{(\autoref{thm:iso-lbm-rt})}
  \\
  \cong
  &\textrm{ Theories of \U-quotients}
  &\text{(\autoref{prop:iso-thUquot-thLBM})}
  \\
  \cong
  &\textrm{ Basic cotheories of regular languages}
  &\text{(Duality)}
  \\
  \cong
  &\textrm{ Basic varieties of regular languages}
  &\text{(\autoref{thm:iso-thlang-varlang})}\tag*{\qed}
  \end{align*}
Spelling out the four isomorphisms in the proof, from top to bottom we transform between the following collections:  
\begin{equation*}
  \begin{array}{cc}
    \text{Collection} & \text{Category} \\
    \hline
    \left.
      \begin{array}{@{}c@{}}
        \V
        \\
        \cong
        \\
        \Th^\V = ( \{ \f[\Sigma] \overset{e}{↠} (M,D) \pipe (M,D)\in \V \} )_{Σ ∈ \A}
      \end{array}
    \quad\right\}
    \rule[23pt]{0mm}{1mm} 
    &
    \LBM 
    \\
    \hspace{-0.8em}\cong\quad
    \\
    (\{Σ^⋄ \overset{e^⋄}{↠} D \pipe e\in \Th^\V \})_{Σ ∈ \A} & \AlgCDL
    \\
    \hspace{-1em}\cong^{\mathsf{op}}
    \\
    \qquad
    \left.
      \begin{array}{@{}c@{}}
        \hspace{4em}(I_Σ \hookrightarrow \Pow(\reg{Σ}))_{Σ ∈ \A}\hspace{-5em}
        \\
        \hspace{7.8em}\cong
        \\
        \hspace{4em}(V_Σ \hookrightarrow \reg{Σ})_{Σ ∈ \A}\hspace{-5em}
      \end{array}
      \qquad
     \rule[23pt]{31mm}{0mm} \right\}
    & \POS
  \end{array}
\end{equation*}
Thus, starting from the top, a pseudovariety $\V$
of lattice bimodules is sent to the basic variety of all regular languages
recognized by some lattice bimodule in
$\V$. Conversely, starting from the bottom, a basic variety $(V_Σ)_{Σ ∈ \A}$ of languages is sent to the pseudovariety of all $\star$-generated
reduced finite lattice bimodules $(M,D)$ such that every language
$L\seq\Sigma^\star$ recognized by $(M,D)$ lies in $V_\Sigma$.

\section{Quantum Finite Automata}\label{sec:applications}

In this section we present a natural example of a basic variety of
regular languages that is not closed under union and intersection and
therefore not captured by any previously known Eilenberg-type
correspondence. It is concerned with languages accepted by \emph{quantum finite automata
(QFA)}. Several different notions of QFA have been proposed and
studied, varying in their expressive power; see
e.g.\ the recent survey paper by Ambainis and
Yakary\i lmaz~\cite{ambainis2015}. Here, we focus on the model of 
\emph{Kondacs-Watrous quantum finite automata (KWQFA)}~\cite{kon-wat},
also known in the literature as \emph{measure-many quantum finite
  automata}. 

 A KWQFA  $M = (Q, \Sigma, T, q_0, Q_\mathsf{acc}, Q_\mathsf{rej},
 Q_\mathsf{non})$ is given by a finite set $Q$ of \emph{basis states},
 an input alphabet $\Sigma$ not containing the end markers $\kappa$
 and $\$$, an initial state $q_0\in Q$ and a partition
 $Q_\mathsf{acc}\dun Q_\mathsf{rej} \dun Q_\mathsf{non}$ of $Q$ into accepting, rejecting and non-halting states. The transitions are specified by a family of unitary linear maps $T_\sigma\colon \H_{Q}\to \H_{Q}$  ($\sigma \in \Sigma\cup \{\kappa,\$\}$) on the complex Hilbert space $\H_{Q}$ with orthonormal basis $Q$. Thus, denoting the basis vectors by $|q\rangle$ ($q\in Q$), every element $|\psi\rangle$ of $\H_Q$ can be uniquely expressed as a linear combination $|\psi\rangle = \sum_{q\in Q} \alpha_q |q\rangle$ with $\alpha_q\in \mathds{C}$. The \emph{states} of $M$ are those $|\psi\rangle\in \H_Q$ with norm $\sum_{q\in Q}|\alpha_q|^2 = 1$. Note that a unitary transformation $T_\sigma$ maps states to states. A \emph{measurement} collapses the state $|\psi\rangle$ to the basis state $|q\rangle$ with probability $|\alpha_q|^2$. 

 Initially, the automaton is in the basis state
 $|q_0\rangle$. An input $w \in
 \Sigma^\star$ is processed by first adding the left
 ($\kappa$) and right ($\$$) end markers. Then, for every successive
 symbol $\sigma$ in $\tilde{w} = \kappa w \$$ the corresponding
 transformation
 $T_{\sigma}$ is applied and a measurement is performed. The automaton
 halts and accepts if the resulting basis state lies in
 $Q_\mathsf{acc}$, halts and rejects if it lies in
 $Q_\mathsf{rej}$, and continues with processing the next input letter
 if it lies in
 $Q_\mathsf{non}$. Thus, if the QFA is in the state $\ket{\psi} =
 \sum_{q \in Q_\mathsf{acc}} \alpha_{q} \ket{q} + \sum_{q \in
   Q_\mathsf{rej}} \beta_{q}\ket{q} + \sum_{q\in Q_\mathsf{non}}
 \gamma_{q}
 \ket{q}$ after reading the current input symbol but before making the
 measurement, it accepts with probability $\sum_{q \in Q_\mathsf{acc}}
 |\alpha_{q}|^2$, rejects with probability $\sum_{q \in Q_\mathsf{rej}}
 |\beta_{q}|^2$ and continues processing the input with probability
 $\sum_{q \in Q_\mathsf{non}}
 |\gamma_{q}|^2$. This yields an overall probability $p\in
 [0,1]$ that the input word
 $w$ is accepted, i.e.~that at any stage of the computation the
 automaton reaches a state in $Q_\mathsf{acc}$.
 
 We say that $M$ \emph{accepts} the language $L \seq \Sigma^{\star}$
 (with bounded error) if there exists a real number $p > 1/2$ such
 that $M$ accepts every word in $L$ with probability $\geq p$ and
 rejects every word not in $L$ with probability $\geq p$. The class of
 languages accepted by KWQFA is denoted by $\RMM$. It is known to be a
 proper subclass of the class of all regular languages; for instance,
 $\{a,b\}^\star a \not\in \RMM$~\cite[Proposition~7]{kon-wat}. Subsequent
 work has identified certain ``forbidden configurations'' in the
 minimal deterministic finite automaton of a regular language making
 it unrecognizable by a KWQFA~\cite{akv2000,bro-pip1999}. In this way,
 it was shown that $\RMM$ is not closed under union and
 intersection~\cite[Corollary~3.2]{akv2000}. However, $\RMM$ is closed
 under preimages of monoid homomorphisms and
 derivatives~\cite[Theorem~4.1]{bro-pip1999} and thus forms a basic
 variety of regular languages.

 The questions whether $\RMM$ is decidable and whether it has an algebraic
 characterization remain open problems in the theory of quantum automata~\cite{abgkmt04}.
 Our Basic Variety Theorem provides
 strong evidence that such a characterization must exist: it asserts
 that $\RMM$ corresponds to a pseudovariety of reduced lattice
 bimodules, which by \autoref{thm:iso-lbm-rt} admits an (abstract form
 of) equational presentation. We expect that the latter can be turned
 into a more concrete form using \emph{profinite equations} over
 free lattice bimodules $\f[\Sigma]$, analogous to
 Reiterman's~\cite{Reiterman82} description of pseudovarieties of
 finite monoids in terms of profinite equations over free monoids
 $\Sigma^\star$. A concrete profinite axiomatization of the
 pseudovariety induced by $\RMM$ might pave the way towards the
 decidability of that class: deciding whether a given regular language
 lies in $\RMM$ reduces to checking whether its syntactic lattice
 bimodule satisfies the equational axioms.



\section{Conclusion and Future Work}
We have introduced a new two-sorted algebraic structure, lattice
bimodules, for the recognition of regular languages. Our main result
is a new Eilenberg-type correspondence between basic varieties of
regular languages, which need not be closed under set-theoretic
boolean operations, and pseudovarieties of reduced lattice bimodules.
The proof is guided by the recent category-theoretic approach to
algebraic language theory and makes use of the duality between
algebraic completely distributive lattices and posets.

An immediate next step to unleash the full power of our new variety theorem is
to establish a Reiterman-type theorem for lattice bimodules leading to a description of pseudovarieties of lattice bimodules in terms of profinite
equations. The recent categorical account of (profinite) equational theories~\cite{camu16,mu19} should provide inspiration in this direction. This may lead to new results on the decidability of
basic varieties of regular languages, e.g.\ language classes recognized by different models of
reversible automata~\cite{gol-pin} or quantum automata (cf.~\autoref{sec:applications}).

Furthermore, several generalizations of our work are conceivable. The
most obvious one is to replace the duality $\AlgCDL\simeq^\op \POS$ by
an abstract dual equivalence $\mathscr{A} \simeq^\op \mathscr{B}$
between suitable categories $\mathscr{A}$ and $\mathscr{B}$, and to
consider the recognition of languages by $\mathscr{A}$-bimodules. We
anticipate that this minor generalization already recovers results
closely related to the original Eilenberg theorem for $\mathscr{A}$
being the category of sets, and to Pol\'ak's variety theorem for
idempotent semirings for $\mathscr{A}$ being the category of complete
semilattices. In an orthogonal direction, the monoid action on the
algebra may be generalized to the action of a monad $\mathbf{T}$ on
the category of sets, but the dependence between the monad
$\mathbf{T}$ and the category $\mathscr{A}$ is not obious and remains
to be investigated.


\bibliographystyle{splncs04}
\bibliography{bibliography-lncs}

\clearpage\appendix

\section*{Appendix}
This appendix provides full proofs and additional details for all our results.

\section{Details for \autoref{sec:lattice-bimodules}}

\begin{lemma}\label{lem:factorization-system}
  The category \LBM has the factorization system of (sortwise) surjective and  injective morphisms. More precisely, every lattice bimodule homomorphism $h$ factorizes as $h=m\cdot e$, where $e$ is a surjective and $m$ is an injective homomorphism, and for every commutative square
\begin{equation*}
  \begin{tikzcd}
    (M_1, D_1)
    \arrow[r, "e", two heads]
    \arrow[d, ""]
    &
    (M_2, D_2)
    \arrow[ld, "d"', dashed]
    \arrow[d, ""]
    \\
    (M_3, D_3)
    \arrow[r, "m", tail]
    &
    (M_4, D_4)
  \end{tikzcd}
\end{equation*}
with $e$ surjective and $m$ injective, there exists a unique \emph{diagonal fill-in} $d$ making both triangles commute.
\end{lemma}
While this lemma is not difficult to prove directly, it also follows immediately from the fact that the category $\LBM$ forms a variety of (infinitary) algebras and is thus monadic over the product category $\Set\times \Set$~\cite{manes1976}. This implies that $\LBM$ inherits the (surjective, injective) factorization system of $\Set\times \Set$.

\begin{notation}\label{not:quo}
 Let $\mathbf{Quo}_{f}^{\C}(X)$ denote the set of all finite quotients (represented by surjective morphisms) of an object $X$ in the category $\C\in \{ \Set,\CDL,\LBM \}$. If \C is clear from context, we may omit the superscript. We  equip $\mathbf{Quo}_{f}^\C(X)$ with the order $e ≤ e'$ if $e$ factorizes through $e'$, i.e.~$e = q · e'$ for some $q$. This makes $\mathbf{Quo}_{f}^\C(X)$ a poset if we identify isomorphic quotients.
\end{notation}

\begin{rem}\label{rem:homtheorem}
Quotients in $\C\in \{ \Set,\CDL,\LBM \}$ satisfy the \emph{homomorphism theorem}:
given two quotients $e, e'$ of the same object, we have $e\leq e'$ if and only if the kernel of $e'$ of contained in the kernel of $e$; that is, for each $x,y$ in the domain of $e'$,
\[ e'(x)=e'(y) \quad\text{implies}\quad e(x)=e(y). \]
\end{rem}

\bigskip\noindent\textbf{Proof of \autoref{prop:lbm-free}}\\
  \newcommand{\he}{\ensuremath{\hat{η}}\xspace}
  \newcommand{\unit}{\ensuremath{(η^⭑, η^⋄)}\xspace}
Let $h_0=(h_0^\star,h_0^\diamond)\colon (\Sigma,\Gamma)\to (M,D)$ be a two-sorted map into a lattice bimodule $(M,D)$. We need to show that there exists a unique $\LBM$-morphism $h\colon (\Sigma,\, \fcdl(\Sigma^\star+\Sigma^\star\times \Gamma\times \Sigma^\star)\to (M,D)$ satisfying $h\cdot \eta = h_0$.

\begin{enumerate}
\item \emph{Existence.} Let $h^\star\colon \Sigma^\star\to M$ be the unique monoid morphism with $h^\star(a)=h_0^\star(a)$ for each $a\in \Sigma$. Moreover, let $h^\diamond\colon \fcdl(\Sigma^\star+\Sigma^\star\times \Gamma\times \Sigma^\star) \to D$ be the unique $\CDL$-morphism with
\begin{equation}\label{eq:heqs} h^\diamond(w)=h^\star(w) \quad\text{and}\quad h^\diamond(v,z,w)= h^\star(v) \tr h_0^\diamond(z) \tl h^\star(w) \end{equation}
for $v,w\in \Sigma^\star$ and $z\in \Gamma$. An easy verification shows that $h=(h^\star,h^\diamond)$ is an $\LBM$-morphism from $(\Sigma^\star,\, \fcdl(\Sigma^\star+\Sigma^\star\times \Gamma\times \Sigma^\star)$ into $(M,D)$ extending $h_0$.
\item \emph{Uniqueness.} Let $h=(h^\star,h^\diamond)$ be any $\LBM$-morphism from  $(\Sigma^\star,\, \fcdl(\Sigma^\star+\Sigma^\star\times \Gamma\times \Sigma^\star)$ into $(M,D)$ extending $h_0$. Then $h^\star(a)=h_0^\star(a)$ for all $a\in \Sigma$ and thus $h^\star$ is the unique monoid morphism extending $h_0$. Moreover, from the fact that $h$ is a homomorphism of lattice bimodules it follows that the equations \eqref{eq:heqs} hold for all $v,w\in \Sigma^\star$ and $z\in \Gamma^\star$, which proves that $h^\diamond$ is uniquely determined by $h_0^\diamond$ and $h^\star$. \qed
\end{enumerate}

\begin{lemma}\label{lemma-projective}
 Free lattice bimodules are projective: for every $\LBM$-morphism $h  \: \f → (M', D')$ and every surjective $\LBM$-morphism $g \: (M, D) ↠ (M', D')$ there exists an $\LBM$-morphism $f  \: \f → (M, D)$ with $h = g·f$.
\end{lemma}
\proof
 For each $a\in \Sigma$ choose $m_a\in M$ with $g^\star(m_a)=h^\star(a)$, using that $g$ is surjective. By the universal property of $\f[\Sigma]$, there exists a unique homomorphism $f  \: \f → (M, D)$ with $f^\star(a)=m_a$ for all $a\in \Sigma$. Then $h=g\cdot f$ since this holds when precomposed with the universal map $\eta\colon (\Sigma,\emptyset)\to \f[\Sigma]$.

\takeout{
\subsection{Monadicity of Lattice Bimodules}\label{subsec:monadicity}
\autoref{lbm-free} provides an adjunction between the categories $\Set × \Set$ and $\LBM$. Naturally the question arises if this adjunction is well-behaved, i.e.~monadic; it indeed is, which follows from the well-known fact that every (possibly infinitary) equational theory with free algebras induces a monad over sets~\cite{manes1976}. In the following we sketch a proof for the convenience of the reader.

\begin{lemma}\label{lem:lbm-monadic}
  The adjunction induced by the free lattice bimodule from \autoref{lbm-free} is monadic.
\end{lemma}
\begin{proof}
  The proof uses Beck's precise tripleability theorem (PTT); see~\cite[Theorem VI.7.1]{cwm}. Let $(h_1, φ_1), (h_1, φ_2)  \: (M, D) → (M', D')$ be homomorphisms of lattice bimodules that have an absolute coequalizer
\begin{equation}\label{eq:coeq}
    \begin{tikzcd}[column sep = 40]
      (M, D)
      \arrow[r, "{(h_1, φ_1)}", shift left=1]
      \arrow[r, "{(h_2, φ_2)}"', shift right=1]
      &
      (M', D')
      \arrow[r, "{(h, φ)}"]
      &
      (N, E),
    \end{tikzcd}
\end{equation}
in $\Set\times\Set$. We are to show that $(h, φ)$ lifts to a coequalizer of lattice bimodules; that is, there exists a unique lattice bimodule structure on $(N, E)$ such that $(h, \phi)$ is a coequalizer of lattice bimodules. We first define a lattice bimodule structure on $(N, E)$. The operations acting only on one sort can be constructed exactly in the same way as in the proof for single-sorted algebras in~\cite[Theorem VI.8.1]{cwm}, since the argument does not depend on the arity of the operation.  We give constructions for the operations $\tr$ and $ι$ acting on multiple sorts in the same manner: We apply the product functor $- × -\: \Set × \Set → \Set$ to the upper and the projection functor to the lower half of diagram~\eqref{eq:coeq} to obtain the following diagram in \Set:
\begin{equation}\label{eq:tr}
    \begin{tikzcd}[column sep = 40]
      M × D
      \arrow[r, "{h_1 × φ_1}" , shift left=1]
      \arrow[r, "{h_2 × φ_2}"', shift right=1]
      \arrow[d, "\tr"]
      &
      M' × D'
      \arrow[r, "{h × φ}"]
      \arrow[d, "\tr'"]
      &
      N × E
      \arrow[d, "", dashed]
      \\
      D
      \arrow[r, "φ_1" , shift left=1]
      \arrow[r, "φ_2"', shift right=1]
      &
      D'
      \arrow[r, "φ"]
      &
      E
    \end{tikzcd}
  \end{equation}
Since  $((h, φ), (N, E))$ is an absolute coequalizer, the upper row of the diagram is a coequalizer. Moreover, we have
 \begin{align*}
  φ·\tr'·(h_1×φ_1) = φ·φ_1·\tr = φ·φ_2·\tr = φ·\tr'·(h_1×φ_1),
 \end{align*}
and thus there is a unique map  $N\times E\to E$ making the right-hand square of the diagram commute. This defines the left action of $M$ on $E$. The definition of the right action $E\times N\to E$ is analogous. The  embedding $N\to E$ can be constructed in the same fashion by applying the projection functor to the top row to get:
\begin{equation*}
    \begin{tikzcd}
      M
      \arrow[r, "h_1" , shift left=1]
      \arrow[r, "h_2"', shift right=1]
      \arrow[d, "ι"]
      &
      M'
      \arrow[r, "h"]
      \arrow[d, "ι'"]
      &
      N
      \arrow[d, "", dashed]
      \\
      D
      \arrow[r, "φ_1" , shift left=1]
      \arrow[r, "φ_2"', shift right=1]
      &
      D'
      \arrow[r, "φ"]
      &
      E
    \end{tikzcd}
  \end{equation*}

  Next we show that $(N, E)$ is indeed a lattice bimodule, i.e.~it satisfies the appropriate equations. We give an example only for the equation $n \tr ⋀_{i ∈ I}e_i = ⋀_{i ∈ I} (n \tr e_i)$ as for other equations one may proceed in exactly the same manner. Since $(h, φ)$ is a coequalizer in $\Set^2$, both $h$ and $φ$ are surjective maps. Choose preimages $m'$ with $h(m') = n$ and $d_i'$ with $φ(d_i') = e_i$. Then, since diagram~\eqref{eq:tr} commutes and $(M', D')$ is a lattice bimodule, we compute
  \begin{align*}
    n \tr ⋀_{i ∈ I}e_i   &= h(m') \tr ⋀_{i ∈ I}φ(d_i') \\
    &= φ(m' \tr ⋀_{i ∈ I}d_i') \\
    &= φ(⋀_{i ∈ I}m' \tr d_i')\\
    &= ⋀_{i ∈ I} (h(m') \tr φ(d_i'))\\
    &= ⋀_{i ∈ I} (n \tr e_i).
  \end{align*}

The remaining proof that $(h, φ)$ is indeed a coequalizer of lattice bimodules is the same as for single-sorted finitary algebras~\cite[Theorem VI.8.1]{cwm}.
\qed\end{proof}

As a consequence of \autoref{lem:lbm-monadic} we get the following corollary that allows us to view lattice bimodules not only as algebras over $\Set \times \Set$ but also over $\Set \times \CDL$.

\begin{corollary}\label{adjunction-split}
  The monadic adjunction induced between $\LBM$ and $\Set\times \Set$ splits into two monadic adjunctions $F\dashv U$ and $L\dashv R$ as shown in the diagram below, where $U$ and $R$ are the forgetful functors.
\begin{equation*}
  \begin{tikzcd}
    &
    &
    \LBM
    \arrow[dd, "R", shift left=1]
    \arrow[dd, "L"', leftarrow, shift right=1]
    \arrow[ddll, "", shift left=1]
    \arrow[ddll, ""', leftarrow, shift right=1]
    \\
    &
    &
    \\
    \Set×\Set
    \arrow[rr, "F", shift left=1]
    \arrow[rr, "U"', leftarrow, shift right=1]
    &
    &
    \Set×\CDL
  \end{tikzcd}
\end{equation*}
\end{corollary}
\begin{proof}
\CDL is monadic over \Set~\cite{ped-woo}, so $\Set ×\CDL$ is monadic over $\Set×\Set$. Since also $\LBM$ is monadic over $\Set\times\Set$ by \autoref{lem:lbm-monadic}, the adjoint lifting theorem~\cite[Corollary 4.5.7]{borceux1994_2} shows that $\LBM$ is monadic over $\Set\times \CDL$.
\qed\end{proof}

With these preparations at hand, we ready to give the
\begin{proof}[Proof of \autoref{lem:factorization-system}]
  The proof uses (a simplified version of)~\cite[Prop.~20.24]{AdamekEA09} :
  \begin{quotation}
Let $\C$ be a category with a factorization system $(\mathcal{E}, \mathcal{M})$ and $\mathbf{T}$ be a monad on $\C$ that preserves $E$, that is, $e \in \mathcal{E}$ implies $\mathbf{T} e \in \mathcal{E}$. Then the category $\cat{Alg(\mathbf{T})}$ of monadic $\mathbf{T}$-algebras has the factorization system of $\mathcal{E}$-carried and $\mathcal{M}$-carried $\mathbf{T}$-homomorphisms.
\end{quotation}
We apply this result to the category $\C=\Set×\Set$ with the factorization system of pairwise surjective/injective morphisms. Note that since a map is surjective if and only if it has a right inverse, every monad on $\Set\times\Set$ preserves surjections, and thus satisfies the conditions of the proposition. In particular, for the monad $\mathbf{T}$ corresponding to the adjunction between $\LBM$ and $\Set\times \Set$ we obtain the desired factorization system for $\LBM \cong \cat{Alg(\mathbf{T})}$.
\qed\end{proof}

\begin{notation}
  In the notation of \autoref{lbm-free}, the free lattice bimodule over a pair of sets $(A, B)$ is given by $(A^⭑, A_B^⋄)$; the notation is inspired by Klíma and Polák 
 ~\cite{kli-pol}. For the case that $B = ∅$ and therefore $A_∅^⋄ = \textrm{FCDL}(A^⭑ + A^⭑ × ∅ × A^⭑) \cong \textrm{FCDL}(A^⭑)$ we drop the subscript $∅$ in $A_∅^⋄$ and just write  \f[A] for the free lattice bimodule over $(A, ∅)$. As the second component is empty we also call \f[A] the free lattice bimodule over the set $A$. We extend this notation to morphisms: For a lattice bimodule homomorphism $f \: (M, D) → (M', D')$ we refer to its first component as $f^⭑  \: M → M'$ and to its second component as $f^⋄  \: D → D'$. Note that we may also (uniquely) extend any monoid homomorphism $h  \: A^\star → B^\star$ to a lattice bimodule homomorphism $(h, \phi) \: \f[A] →  \f[B]$ by simply defining $\phi(  ⋁_{i ∈ I} ⋀_{k ∈ K_i} \iota(a_{ik})) =   ⋁_{i ∈ I} ⋀_{k ∈ K_i} \iota(h(a_{ik}))$ and so lattice bimodule homomorphisms are in bijection with monoid homomorphisms $A^⭑ → B^⭑$.
\end{notation}
}
\begin{lemma}\label{lem:lbm-internal-monoid}
  For every lattice bimodule $(M, D)$ the operation $ι\colon M\to D$ induces
  a monoid congruence $≡_ι$ on $M$ given by \[m\equiv_\iota n \quad\text{iff}\quad \iota(m)=\iota(n).\] Thus, $ι[M]$ carries a monoid structure with $ι[M] \cong M/\mathord{≡_ι}$.
\end{lemma}

\begin{proof}
Clearly $\equiv_\iota$ is an equivalence relation. For $m_1 ≡_ι m_2, n_1 ≡_ι n_2$ we have
\begin{align*}
\iota(m_1n_1) &= \iota(m_1)\tr n_1 \\
&= \iota(m_2)\tr n_1 \\
&= \iota(m_2n_1) \\
&= m_2 \tl \iota(n_1) \\
&= m_2 \tl \iota(n_2) \\
&= \iota(m_2n_2)
\end{align*}
and thus $m_1n_1\equiv_\iota m_2n_2$, showing that $\equiv_\iota$ is a monoid congruence.
\qed\end{proof}

\bigskip\noindent\textbf{Proof of \autoref{lem:reduced-vs-embedded}}
\begin{enumerate}
\item Let $h \: \f ↠ (M, D)$ be surjective. Given $d \in D$, choose an element $W = ⋁_{j\in J}⋀_{k\in K_j}\iota(w_{jk}) ∈ Σ^⋄$, where $w_{jk}\in \Sigma^\star$, such that $h(W) = d$. Then
  \begin{align*}
    d = h(W) = h\big( \distrib{ι(w_{jk})}\big) 
    = \distrib{h(ι(w_{jk}))} = \distrib{ ι(h(w_{jk}))},
  \end{align*}
  proving that $(M, D)$ is ⭑-generated.

  Conversely, suppose that $(M, D)$ is ⭑-generated. Choose $\Sigma=M$
  and let $h\colon \f[\Sigma]\to (M,D)$ be the unique lattice bimodule
  morphism with $h^\star(a)=a$ for every $a\in \Sigma$. Since $D$ is
  ⭑-generated, for each $d \in D$ we have
  \[
    d = \distrib{ι(m_{jk})}.
  \]
for some $m_{jk}\in M$.
Since $h^\star$ is surjective, there exist $w_{jk}\in \Sigma^\star$ with $h^\star(w_{jk})=m_{jk}$. It follows that
    \begin{align*}
      d &= \distrib{ι(h^\star(w_{jk}))}\\
        &= \distrib{h^\diamond(ι(w_{jk}))} \\
        &= h^\diamond(\distrib{ι(w_{jk})})
    \end{align*}
    and so both components of $h$ are surjective.
\item   Let $(M, D)$ be an ⭑-embedded lattice bimodule and let
    $h \: (M, D) ↠ (M', D')$
  be a quotient with $h^\diamond$ a $\CDL$-isomorphism. Then $h^\diamond$ is injective and thus $h^\diamond·ι = ι'·h^\star$ is injective as well. Hence $h^\star$ is injective, and so $h$ is an isomorphism
\item  Let $(M, D)$ be a ⭑-generated reduced lattice bimodule. Since $(M, D)$ is ⭑-ge\-ne\-rated, we can define a left action of the monoid $ι[M]$ on $D$ by
  \begin{align*}
    [m]_ι \tr \distrib{ι(m_{jk})} := \distrib{m \tr ι(m_{jk})}
  \end{align*}
  where $[m]_ι$ is the equivalence class of $m$ under $≡_ι$, see~\autoref{lem:lbm-internal-monoid}. 
  It is well defined since for $m ≡_ι n$ we have
  \begin{align*}
    [m]_ι \tr \distrib{ι(m_{jk})} &= \distrib{m \tr ι(m_{jk})}\\
    &= \distrib{ι(m·m_{jk})}\\
                                  &= \distrib{ι(m) \tl m_{jk}}\\
                                  &= \distrib{ι(n) \tl m_{jk}}\\
                                  &= [n]_ι \tr \distrib{ι(m_{jk})}.\\
  \end{align*}
  Thus $(ι[M],D)$ carries the structure of a lattice bimodule with $\tr$ defined as above, $\tl$ defined symmetrically, and the unary operation $\iota[M]\to D$ given by $[m]_\iota \to \iota(m)$. Moreover, letting $\iota'\colon M\epito \iota[M]$ denote the codomain restriction of $\iota$, we see that
$
    (ι', id) \: (M, D) ↠ (ι[M], D)
  $
  is a lattice bimodule homomorphism. Since $(M,D)$ is reduced and $id$ is an isomorphism, we conclude that $ι'$ is an isomorphism. This implies that $\iota$ is injective, so $(M,D)$ is ⭑-embedded.\qed
\end{enumerate}

\begin{lemma}\label{lem:star-gen-closed}
    Finite products and quotients of finite ⭑-generated lattice bimodules are again ⭑-generated.
\end{lemma}
\begin{proof}
The statement for quotients follows immediately from~\autoref{lem:reduced-vs-embedded}(1). As for products,
  it suffices to prove the statement for binary products since the $n$-ary case follows by iteration and the empty product is the trivial lattice bialgebra $(1,1)$, which is obviously $\star$-generated. Let $(M, D), (M', D')$ be ⭑-generated. To show that $(M,D)\times (M',D')=(M\times M',D\times D')$ is $\star$-generated, let $(d,d')\in D\times D'$; we need to show $(d, d')$ to be generated by elements of $M \times M'$. By hypothesis, there exist elements $m_{lk}\in M$ and $m'_{rk}\in M'$ such that
    \begin{align*}
        d = \bigvee_{l=1}^n \bigwedge_{k=1}^{n_l} m_{lk}
    \quad\text{and}\quad d' = \bigvee_{r=1}^m \bigwedge_{k=1}^{m_r} m'_{rk}.
    \end{align*}
 for some natural numbers $n,m,n_l,m_r$. (For notational convencience, we identify elements $m\in M$ and $m'\in M'$ with their images $\iota(m)\in D$ and $\iota(m')\in D'$.) Let
    \begin{align*}
o = \max\{n_l, m_r \mathrel{|} l=1 \ldots n, r=1 \ldots m\}.
    \end{align*}
    Using idempotence, extend the conjunctions by the first factor so each conjunction is the same size.
    \begin{align*}
      ⋀_{k=1}^{n_l} m_{lk} &↦
      \big(⋀_{i=1}^{n_l} m_{lk}\big) ∧ \big(⋀_{i=1}^{o-n_l} m_{l1}\big) =: T_l\\
      ⋀_{k=1}^{m_r} m_{rk}' &↦
      \big(⋀_{i=1}^{m_r} m_{rk}'\big) ∧ \big(⋀_{i=1}^{o-m_r} m_{r1}'\big) =: T'_r,
    \end{align*}
so $d = ⋁_{l_1}^n T_l =: L$ and $d' = ⋁_{r=1}^m T'_r$.
    Without loss of generality we can assume $m ≤ n$.
    Then, using idempotence laws again, extend $d'$ to having as many disjuncts as $d$ to get
    \begin{align*}
d = ⋁_{r=1}^m T'_r = \big(⋁_{r=1}^m T'_r\big) \lor \big(\bigvee_{r=1}^{n-m} T'_1\big) =: R.
    \end{align*}
    Now the terms $L$ and $R$ have an equal number of conjuncts and disjuncts so there exist $n_{αβ} ∈ M, n'_{αβ} ∈ M'$ with
    \begin{align*}
      \textstyle
(d, d') = (⋁_α⋀_β n_{αβ}, ⋁_α⋀_β n'_{αβ}) = ⋁_α⋀_β (n_{αβ}, n'_{αβ})
\end{align*}
which proves $D × D'$ to be ⭑-generated.
\qed\end{proof}

\begin{lemma}\label{lem:inj-lbm-sub-prod}
  Subbimodules and products of ⭑-embedded lattice bimodules are ⭑-embedded.
\end{lemma}
\begin{proof}
  Restrictions and products of injective functions are injective.
\qed\end{proof}


\begin{expl} In general, subbimodules of $\star$-generated lattice bimodules are not $\star$-generated, and quotient bimodules of reduced lattice bimodules are not reduced. To see this, consider the lattice bimodule $(\Int/2 \Int,D)$ where $\Int/2 \Int$ is the additive group of integers modulo $2$, and $D = \{\bot, \top, \bar{0}, \bar{1}\}$ is the diamond lattice induced by the order $\bot ≤ \bar{i} ≤ \top, i \in \Int/2 \Int$. The operation $\iota\: \Int/2 \Int \monoto D$ is the obvious injection; this determines $\tr$ and $\tl$ uniquely. Then $(\Int/2 \Int,D)$ is $\star$-generated and $\star$-embedded and thus reduced by \autoref{lem:reduced-vs-embedded}. However:
\begin{enumerate}
\item The subbimodule $(\{\bar{0}\},D)\monoto (\mathbb{Z}/2\mathbb{Z},D)$ is not $\star$-generated.
\item The quotient bimodule $(id,!)\colon (\mathbb{Z}/2\mathbb{Z},D)\epito (\mathbb{Z}/2\mathbb{Z},1)$ is not reduced since it is not ⭑-embedded even though it is ⭑-generated.
\end{enumerate}
\end{expl}

\begin{defn}
A \emph{lattice bimodule congruence} $\equiv$ on a lattice bimodule $(M, D)$ is a pair
$(\e{M}\,\subseteq M × M, \e{D}\,\subseteq D × D)$
such that $\e{M}$ is a monoid congruence on $M$, $\e{D}$ is a complete lattice congruence on $D$ and the operations between the two sets preserve the congruences, that is, for all $m, m' ∈ M $ and  $d, d' ∈ D$,
\begin{itemize}
    \item $m \e{M} m' \text{ implies } ι(m) \e{D} ι(m')$;
    \item $m \e{M} m' \text{ and } d \e{D} d' \text{ implies }
            m \triangleright d \e{D} m' \triangleright d' \text{ and }
            d \triangleleft m \e{D} d' \triangleleft m'$.
          \end{itemize}
        \end{defn}

\begin{rem}\label{rem:quotients-vs-congruences}
Lattice bimodule congruences on $(M,D)$ correspond uniquely to
quotient lattice bimodules of $(M,D)$. More precisely, a quotient
\[
  e=(e^\star,e^\diamond)\colon (M,D)\epito (M',D')
\]
in $\Set\times \CDL$ carries a quotient lattice bimodule (i.e.~there exists a lattice bimodule structure on $(M',D')$ making $e$ a homomorphism of lattice bimodules) if and only if its kernel relation $(\equiv_M,\equiv_D)$, defined by
\[ m\equiv_M m \quad\text{iff}\quad e^\star(m)=e^\star(m') \qquad\text{and}\qquad d\equiv_D d'\quad \text{iff}\quad e^\diamond(d)=e^\diamond(d'),\]
forms a lattice bimodule congruence. This follows immediately from the homomorphism theorem (see \autoref{rem:homtheorem}).
\end{rem}


\bigskip\noindent\textbf{Proof of \autoref{L:up}}\\
    Let $\mathord{\equiv} = (\e{M}, \e{D})$ be a pair of an equivalence relation on $M$ and a CDL congruence on $D$. By the equivalence of quotients and congruences, see~\autoref{rem:quotients-vs-congruences}, and the homomorphism theorem, it suffices to show that $\equiv$ is a lattice bimodule congruence iff it is stable under the unary operations in $\mathbb{U}$. The latter means that for all $u\: S \to T$ in $\U$, where $S,T \in
\{M,D\}$, and all $a,b\in S$ with $a \equiv_S b$ we have $u(a) \equiv_T u(b)$.
   
 Clearly every \LBM-congruence is stable under $\U$. Conversely, suppose that $\equiv$ is stable under $\mathbb{U}$ and that $m \e{M} m', n \e{M} n', d \e{D} d'$. Then
    \begin{itemize}
        \item $m · n = (m \, ·)(n) \e{M} (m \, ·)(n') = m · n' = (· \, n')(m) \e{M} (· \, n')(m') = m' · n'$ because $(m\, \cdot), (\cdot\, n')\in \U$. Thus, $\e{M}$ is a monoid congruence.
        \item $ι(m) \e{D} ι(m')$ because $\iota\in \U$.
        \item $m \tr d = (m \, \tr)(d) \e{D} (m \, \tr)(d') = m \tr d' = (\tr \, d')(m) \e{D} (\tr \, d')(m') = m' \tr d'$ because $(m\,\tr), (\tr\, d')\in \U$.
        \item $d \tl m \e{D} d' \tl m'$, analogously.
    \end{itemize}
    Thus, $\equiv$ is an \LBM-congruence.
\qed

\section{Details for \autoref{sec:pseudovarieties}}

\bigskip\noindent\textbf{Remark on \autoref{D:pseudovar}}\\
Notice that in \ref{D:pseudovar:2} the bimodule $(M',D')$ is necessarily
$\ast$-generated by \autoref{lem:star-gen-closed}. Similarly, in~\ref{D:pseudovar:3} the bimodule $(M,D)$ is necessarily reduced
by %
\autoref{lem:inj-lbm-sub-prod} and~\ref{lem:reduced-vs-embedded}.

\begin{lemma}\label{reduced-theory-factor}
  Let $\Th$ be a theory of lattice bimodules and let $(M, D)$ be a ⭑-generated reduced finite lattice bimodule. Then the following are equivalent:
  \begin{enumerate}
  \item\label{reduced-theory-factor:1}
    There exists $h ∈ \locT$ with codomain $(M, D)$ for some $Σ\in \Setf$.
  \item\label{reduced-theory-factor:2}
    Every lattice bimodule homomorphism $f  \: \f[Δ] → (M, D)$ with $\Delta\in \Setf$ factorizes through some element of  $\locT[Δ]$.
  \end{enumerate}
\end{lemma}

\begin{proof}
  For \ref{reduced-theory-factor:1}$\Rightarrow$\ref{reduced-theory-factor:2}, let  $f \: \f[Δ] → (M, D)$. By hypothesis there exists an alphabet Σ and a quotient $h \: \f ↠ (M, D)$ in $\Th_Σ$. Using \autoref{lemma-projective} we can choose a morphism of lattice bimodules $g  \: \f[Δ] → \f[Σ]$ with $f = h · g$. Since $\Th$ is a theory, $f = h·g$ factorizes through some $\bar{h} ∈ \Th_Δ$.
\begin{equation*}
  \begin{tikzcd}
    \f[Δ]
    \arrow[r, "g"]
    \arrow[rd, "f"]
    \arrow[d, "\bar{h}"', two heads]
    & \f[Σ]
    \arrow[d, "h" , two heads]
    \\
    (M', D')
    \arrow[r, "\bar{g}"]
    &
    (M, D)
  \end{tikzcd}
\end{equation*}
  For \ref{reduced-theory-factor:2}$\Rightarrow$\ref{reduced-theory-factor:1}, suppose that  $(M,D)$ satisfies~\ref{reduced-theory-factor:2}. Since $(M, D)$ is ⭑-generated, there exists a surjective homomorphism $f  \:  \f[Δ] ↠ (M, D)$ for some $Δ\in \Setf$ by~\autoref{lem:reduced-vs-embedded}(1); in fact, the proof of that lemma shows that one can choose $\Delta=M$. By assumption, $f$ factors through some $h ∈ \Th_Δ$, i.e.~$f ≤ h$. Since $\Th_Δ$ is downwards closed, we conclude $f ∈ \Th_Δ$.
\qed\end{proof}
Recall from \autoref{not:vt-tv} the class $\mcpower{V}{T}$ associated to a theory $\Th$. The elements of $\mcpower{V}{T}$ are those $\star$-generated reduced finite lattice bimodules satisfying the  equivalent conditions of \autoref{reduced-theory-factor}.
\begin{lemma}
  If $\Th$ is a theory of lattice bimodules, then $\mcpower{V}{T}$  is a pseudovariety of lattice bimodules.
\end{lemma}

\begin{proof}
  Let $\Th$ be a theory. The class $\mcpower{V}{T}$ is closed under reduced quotients because all \locT are downwards closed. To show that $\mcpower{V}{T}$ is closed under ⭑-generated subbimodules of finite products, suppose that $ m \: (M, D) ↣ ∏_{i∈I}(M_i, D_i)$ is such a subbimodule with $(M_i, D_i) ∈ \mcpower{V}{T}$, $I$ finite. We prove $(M, D) ∈ \mcpower{V}{T}$ by condition \ref{reduced-theory-factor:2} of \autoref{reduced-theory-factor}, viz.\ that any $h  \: \f[Δ] → (M, D)$  factors through $\locT[Δ]$. Let $p_i\: ∏_{i∈I}(M_i, D_i)\to (M_i,D_i)$ denote the product projections. Then, since $(M_i,D_i)\in \V^\Th$, the homomorphisms $p_i · m · h$ each factor through some $f_i$ in $\Th_Δ$ via $k_i  \: (\overline{M_i}, \overline{D_i}) → (M_i, D_i)$. Since $\Th_Δ$ is a local pseudovariety and $I$ is finite,  the $f_i$ have an upper bound $f\in \Th_\Delta$, i.e.~$f_i=l_i\cdot f$ for some $l_i$. Then the diagonal fill-in property, applied to the left square in the commutative diagram below, yields $g \: (\overline{M}, \overline{D}) → (M, D)$ with $h=g\cdot f$. This proves $(M,D)\in \V^\Th$. \qed
  \begin{equation*}
    \begin{tikzcd}
      (Δ^⭑, Δ^⋄)
      \arrow[d, "h"']
      \arrow[->>,shiftarr = {yshift=20}]{rr}{f_i}
      \arrow[r , "f", two heads]
      &
      (\overline{M}, \overline{D})
      \arrow[r, "l_i", two heads]
      \arrow[d, "⟨k_i · l_i⟩"]
      \arrow[ld, "g"', dashed]
      &
      (\overline{M_i}, \overline{D_i})
      \arrow[d, "k_i"]
       \\
      (M, D)
      \arrow[r, "m", tail]
      &
      ∏_{i∈I}(M_i, D_i)
      \arrow[r, "p_i"]
      &
      (M_i, D_i)
    \end{tikzcd}
  \end{equation*}
\end{proof}

\begin{lemma}
  For any pseudovariety of reduced lattice bimodules $\V$, the family $\mcpower{T}{V}$ (see \autoref{not:vt-tv}) is a theory of lattice bimodules.
\end{lemma}
\proof 
We first show each $\mcpower{T}{V}_Σ$ to be a local pseudovariety. It is downwards closed since $\mathcal{V}$ is closed under reduced quotients. To show directness let $e_i \: \f ↠ (M_i, D_i)$, $i=1,2$, be two quotients in $\mcpower{T}{V}_Σ$. The image $\langle e_1, e_2\rangle[\f] ⊆ (M_1, D_1) × (M_2, D_2)$ of \f under $\langle e_1 , e_2 \rangle$  is ⭑-generated by \autoref{lem:reduced-vs-embedded}(1) and reduced by \autoref{lem:inj-lbm-sub-prod}. Since $\mathcal{V}$ is a pseudovariety, is follows that the lattice bimodule $\langle e_1, e_2\rangle[\f]$ lies in $\mathcal{V}$. Therefore, the codomain restriction of $\langle e_1, e_2\rangle$ to its image $\langle e_1, e_2\rangle[\f]$ is an element of $\Th_\Sigma^\V$. It is an upper bound for both $e_1$ and $e_2$, so $\mcpower{T}{V}$ is directed.

To confirm that $\mcpower{T}{V}$ is a theory, let $e ∈ \mcpower{T}{V}_Σ$ with codomain $(M, D)$, and $h \: \f[Δ] → \f[Σ]$. Factorize $e·h$ into a surjective lattice bimodule homomorphism $\bar{e}  \: \f[Δ] ↠ (M', D')$ followed by an injective homomorphism $\bar{h}  \: (M', D') → (M, D)$. Then $(M', D')$ is a ⭑-generated subbimodule of $(M, D)$ and therefore itself in $\mathcal{V}$, since $\mathcal{V}$ is a pseudovariety. Thus $\bar{e} ∈ \mcpower{T}{V}_Δ$, as required.
  \begin{equation*}
    \begin{tikzcd}[baseline = (MD.base)]
      \f[Δ]
      \arrow[r, "h"]
      \arrow[d, "\bar{e}"', two heads]
      &
      \f[Σ]
      \arrow[d, "e", two heads]
      \\
      (M', D')
      \arrow[r, "\bar{h}", tail]
      &
      |[alias = MD]|
      (M, D)
    \end{tikzcd}
    \tag*{\qed}
  \end{equation*}

\bigskip\noindent\textbf{Proof of \autoref{thm:iso-lbm-rt}}
\begin{enumerate}
\item\label{thm:iso-lbm-rt:1}  For any pseudovariety $\mathcal{V}$ of lattice bimodules it holds that $\mathcal{V} = \mcpower{V}{T}$ where $\Th := \mcpower{T}{V}$. Indeed, to show $\mathcal{V} ⊆ \mcpower{V}{T}$, let $(M, D) ∈ \mathcal{V}$. Since $(M, D)$ is ⭑-generated, there exists a surjective homomorphism $h \: \f ↠ (M, D)$ for some $\Sigma\in\Setf$. Then $h ∈ \Th_\Sigma$ by the definition of $\Th$, and so $(M, D) ∈ \mcpower{V}{T}$. Conversely, to show $\mcpower{V}{T} ⊆ \mathcal{V}$ suppose that $(M, D) ∈ \mcpower{V}{T}$. Then there exists a $h ∈ \mcpower{T}{V}_Σ$ with codomain $(M, D)$. But then, by definition of $\mcpower{T}{V}$, $(M, D)$ must have been in $\mathcal{V}$.
\item\label{thm:iso-lbm-rt:2}
  For any theory $\Th$ of lattice bimodules it holds that $\Th = \mcpower{T}{V}$, where $\mathcal{V} = \mcpower{V}{T}$.

  To see this, we first show $\Th ⊆ \mcpower{T}{V}$. For $h ∈ \Th_Σ$ with codomain $(M, D)$ we get $(M, D) ∈ \mathcal{V}$, so also $h ∈ \mcpower{T}{V}_Σ$. For the direction $\mcpower{T}{V} ⊆ \Th$ suppose that $h ∈ \mcpower{T}{V}_Σ$. Then its codomain $(M, D)$ is a lattice bimodule in $\mathcal{V}$, so by definition there exists some $e ∈ \locT[Δ]$ with codomain $(M, D)$. By \autoref{lemma-projective} we can choose a $g \: \f[Σ] → \f[Δ]$ with $h = e·g$. Since $\Th$ is a theory, there exist $\bar{e} ∈ \locT[Σ]$ and $\bar{g}$ to make the diagram below commute. Then $h ≤ \bar{e} ∈ \locT$ and therefore $h ∈ \locT$ because $\locT$ is downwards closed.
  \begin{equation*}
    \begin{tikzcd}
      \f
      \arrow[r, "g"]
      \arrow[rd, "h", two heads]
      \arrow[d, "\bar{e}"', two heads]
      &
      \f[Δ]
      \arrow[d, "e", two heads]
      \\
      (M', D')
      \arrow[r, "\bar{g}"]
      &
      (M, D)
    \end{tikzcd}
  \end{equation*}
\item Parts~\ref{thm:iso-lbm-rt:1} and~\ref{thm:iso-lbm-rt:2} show that the maps $\mathcal{V} ↦ \mcpower{T}{V}$ and $\Th ↦ \mcpower{V}{T}$ are mutually inverse bijections. Moreover, by definition both maps are clearly order-preserving, which shows that they define an isomorphism of lattices. \qed
\end{enumerate}
Our next aim is to prove \autoref{prop:iso-thUquot-thLBM}. The key to this result lies in the observation that \U-quotients  of $\Sigma^\diamond$ and reduced quotients of \f are in one-to-one correspondence. This is based on the following construction:

\begin{notation}\label{not:eqphi}
\begin{enumerate}
\item For notational simplicity, for a lattice bimodule homomorphism $h=(h^\star,h^\diamond)$, we sometimes omit the superscripts $(-)^\star$ and $(-)^\diamond$ and denote both components by $h$.
\item For any quotient $e \: Σ^⋄ ↠ D$ in \CDL, we define a pair
$≡_e\,\,= (≡^⭑_e, ≡^⋄_e)$ of equivalence relations on \f as follows: For $x, y\in \Sigma^s$, where $s ∈ \{⭑, ⋄\}$, put
\[x ≡^s_e y \quad\text{iff}\quad  e(u(x)) = e(u(y)) \text{ for every $u ∈ \barU(\Sigma^{s}, \Sigma^{\diamond})$}.\]
Note that $≡^⋄_e$ is a \CDL-congruence because all $u ∈ \barU(\Sigma^⋄,\Sigma^⋄)$ are \CDL-mor\-phisms. Moreover, $≡_e$ is stable under all unary operations in $\U$ since $\barU$ is closed under composition. Thus, \autoref{L:up} (see also \autoref{rem:quotients-vs-congruences}) shows that $\equiv_e$ induces a quotient lattice bimodule of $\f$, denoted by
\begin{equation*}
e_R = (e_R^⭑, e_R^⋄) \: \f ↠ (Σ^⭑/e, Σ^⋄/e).
\end{equation*}
Note that there is no semantic ambiguity in the term $e_R^⋄$. If $e ∈ \CDL$ then it can only be read as $(e_R)^⋄$ and conversely for $e ∈ \LBM$ as $(e^⋄)_R$.
\end{enumerate}
\end{notation}
The key properties of the quotient $e_R$ are established by the next lemma:

\begin{lemma}\label{reduced-quotients-props}
For any $\CDL$-quotient $e \: Σ^⋄ ↠ D$, the following holds true:
  \begin{enumerate}
  \item\label{reduced-quotients-props:1}
    $e_R$ is the smallest lattice bimodule quotient of \f with $e ≤ e^⋄_R$.
  \item\label{reduced-quotients-props:2}
    The lattice bimodule $(Σ^⭑/e, Σ^⋄/e)$ is reduced.
  \item
    If $e$ is a \U-quotient, then $e_R$ is the unique reduced lattice bimodule quotient with $e_R^⋄ = e$.
  \end{enumerate}
\end{lemma}

\begin{proof}
  We first prove the following auxiliary statement $(\#)$:
  \begin{center}
  For any lattice bimodule quotient $h = (h^⭑, h^⋄) \: \f ↠ (M', D')$:\\ Whenever $e ≤ h^⋄$ then also $e_R ≤ h$.
\end{center}
  To see this, let $e ≤ h^⋄$, so $e = g·h^⋄$ for some $g$. To prove that $e_R ≤ h$, we apply the homomorphism theorem: given $x, y ∈ Σ^s$ with $h(x) = h(y)$ we need to show that $e_R(x) = e_R(y)$, that is, $(e·u)(x) = (e·u)(y)$ for every $u ∈ \barU(\Sigma^s,\Sigma^⋄)$. Since $\barU$ is a unary presentation and $h$ is a lattice bimodule quotient there exists a lifting $\bar{u}$ of $u$ along $h$. Then
  \begin{align*}
    e(u(x)) = g(h(u(x))) = g(\bar{u}(h(x)) = g(\bar{u}(h(y)) = g(h(u(y))) = e(u(y)),
  \end{align*}
 as required. Now we proceed to prove the statements from the lemma.
  \begin{enumerate}
  \item
    First, we use the homomorphism theorem to prove $e ≤ e^⋄_R$. Let $x, y ∈ Σ^⋄$ with $e^⋄_R(x) = e^⋄_R(y)$, so $e(u(x)) = e(u(y))$ for any $u ∈ \barU(\Sigma^⋄, \Sigma^⋄)$. In particular $e(x) = e(id(x)) = e(id(y)) = e(y)$ since $id ∈ \barU(\Sigma^⋄,\Sigma^⋄)$. That $e_R$ is the smallest lattice bimodule quotient of \f with $e ≤ e_R^⋄$ follows from $(\#)$.
  \item
    Given a lattice bimodule quotient $g = (g^⭑, g^⋄) \: (Σ^\star/e, Σ^⋄/e) ↠ (M', D')$ with $g^⋄$ an isomorphism in $\CDL$ we need to show that $g$ is an isomorphism in $\LBM$. Since trivially $g·e_R ≤ e_R$, the opposite $e_R ≤ g · e_R$ would suffice to prove $g$ an isomorphism. To do so, we first show $e ≤ g^⋄ · e^⋄_R$ by using the homomorphism theorem: For $x, y ∈ Σ^⋄$ with $g^⋄(e^⋄_R(x)) = g^⋄(e^⋄_R(y))$ we get  $e^⋄_R(x) = e^⋄_R(y)$ because $g^\diamond$ is an isomorphism and thus $e(x) = e(y)$ by part (1) of this lemma. Now to derive  $e_R ≤ g · e_R$ we apply $(\#)$.
  \item
    Suppose that $e$ is a \U-quotient. We prove $e_R^⋄ = e$ by showing they have the same kernel: For $x, y ∈ Σ^⋄$, we have $x ≡_e^⋄ y$ iff $e(x) = e(y)$. The ``only if'' direction follows directly from part (1). Conversely, let $e(x) = e(y)$. For all $u ∈ \barU(\Sigma^⋄, \Sigma^⋄)$ with a lifting $\bar{u}$ along $e$ we compute $e(u(x)) = \bar{u}(e(x)) = \bar{u}(e(y)) = e(u(y))$, so $x ≡_e^⋄ y$ and this proves the ``if'' direction.

    For the uniqueness suppose that $h=(h^⭑, h^⋄) \: \f → (M', D)$ is reduced with $h^⋄ = e$. Then trivially $e ≤ h^⋄$ and therefore $e_R ≤ h$ by $(\#)$. Thus, there exists a homomorphism $g$ with $e_R = g·h$ and since $e = e_R^⋄ = g^⋄ · h^⋄ = g^⋄ · e$ we see that $g^⋄=id$ since $e$ is epi. Since $(M',D)$ is reduced we conclude that $g$ is a isomorphism and thus $e_R$ and $h$ form the same quotient of $\f$.\qed
  \end{enumerate}

\end{proof}


\begin{rem}\label{rem:gen-quot-u-quot}
For any lattice bimodule quotient $h \: \f ↠ (M, D)$ the second component $h^⋄$ is a \U-quotient; this follows immediately from the fact that \barU is a unary presentation of \f.

\end{rem}

\begin{lemma}\label{reduced-quotients-bijection}
  The maps $h ↦ h^⋄$ and $e ↦ e_R$ define an isomorphism between the poset of reduced lattice bimodule quotients of \f and the poset of \U-quotients of $Σ^⋄$.
\end{lemma}

\begin{proof}
    We show the assignments to be mutually inverse. Let $e \: Σ^⋄ ↠ D$ be a \U-quotient. Then \autoref{reduced-quotients-props}.3 shows that $e_R^\diamond = e$. Conversely, if $h \: \f ↠ (M, D)$ is reduced then $h^⋄$ is a \U-quotient, so it follows from \autoref{reduced-quotients-props}.3 that $(h^⋄)_R = h$ since they agree on the second component.
    The map $h ↦ h^⋄$ is clearly monotone. To show that the map $e ↦ e_R$ is monotone take \U-quotients $e, f \: Σ^⋄ ↠ D_i$ with $e ≤ f$. Then $e ≤ f ≤ f^⋄_R$ by \autoref{reduced-quotients-props}.1 and hence $e_R ≤ f_R$, again by \autoref{reduced-quotients-props}.1.
  \qed\end{proof}

%
%

\bigskip\noindent\textbf{Proof of \autoref{prop:iso-thUquot-thLBM}}\\
We show that the isomorphism of \autoref{reduced-quotients-bijection} induces an isomorphism 
between theories of reduced lattice bimodules and theories of $\U$-quotients. Explicitly, this isomorphism maps a theory $\Th$ of reduced lattice bimodules to the theory $\Th^\diamond$ of $\U$-quotients containing all quotients $e^\diamond$ with $e\in \Th$. Its inverse maps a theory $\Th$ of $\U$-quotients to the theory $\Th_R$ of reduced lattice bimodules containing all $e_R$ with $e\in \Th$. Since these maps are clearly mutually inverse, the only thing we need to show is that they are well-defined, i.e. they actually map theories to theories.
\begin{enumerate}
\item Given a theory of reduced lattice bimodules $\Th = \T$ we show that the corresponding family $\Th^\diamond$ is a theory of \U-quotients. Clearly, each $\Th_\Sigma^\diamond$ is an ideal since $\Th_\Sigma$ is an ideal.  Given a \U-quotient $e^\diamond$ that is the second component of some $e  \: \f ↠ (M, D)$ in $\locT$ and $g \: \f[Δ] → \f[Σ]$, choose a reduced lattice bimodule quotient $\overline{e}  \: \f[\Delta] ↠ (M', D') $ in $\locT[Δ]$ and a lifting $\bar{g} \: (M', D') → (M, D)$ with $\bar{g}·\bar{e} = e·g$. Dropping the ⭑-component yields
\begin{equation*}
  \begin{tikzcd}
    Δ^⋄
    \arrow[r, "g^⋄"]
    \arrow[d, "\bar{e}^⋄"', two heads]
    &
    Σ^⋄
    \arrow[d, "e^\diamond", two heads]
    \\
    D'
    \arrow[r, "\bar{g}^⋄"']
    &
    D
  \end{tikzcd}
\end{equation*}
and thus the desired lifting for $e^\diamond·g^\diamond$ along $\bar{e}^\diamond$ in $\Th^\diamond$. This proves that $\Th^\diamond$ is a theory of $\U$-quotients.
\item Given a theory $\Th$ of \U-quotients, we show that $\Th_R$ is a theory of reduced lattice bimodules. Clearly, each $(\Th_R)_\Sigma$ is an ideal since $\Th_\Sigma$ is an ideal. Now let $e \: Σ^⋄ ↠ D$ be a \U-quotient in $\locT$ and $f \: \f[Δ] → \f$. We need to show that $e_R·f$ has a lifting. Apply the lifting property of $\Th$ to $e$ to obtain the following commutative diagram, where $\bar{e}\in \Th_\Delta$ is a lifting of $e·f$:
\begin{equation*}
  \begin{tikzcd}
    \f[Δ]
    \arrow[r, "f"]
    \arrow[d, "\bar{e}_R"', two heads]
    \arrow[shiftarr = {xshift=-30},->>]{dd}[swap]{(!,\bar{e})}
    &
    \f[Σ]
    \arrow[d, "e_R", two heads]
    \arrow[shiftarr = {xshift=30},->>]{dd}{(!,e)}
    \\
    (Δ^⭑/\bar{e}, D')
    \arrow[d, "{(!,id)}"', two heads]
    &
    (Σ^⭑/e, D)
    \arrow[d, "{(!, id)}", two heads]
    \\
    (1, D')
    \arrow[r, "{(!,\hat{f})}"', dotted]
    &
    (1, D)
  \end{tikzcd}
\end{equation*}
What remains is to find the first component $\hat{h}$ of the dashed arrow below:
\begin{equation*}
  \begin{tikzcd}
    \f[Δ]
    \arrow[r, "f"]
    \arrow[d, "\bar{e}_R"', two heads]
    \arrow[shiftarr = {xshift=-30},->>]{dd}[swap]{(!,\bar{e})}
    &
    \f[Σ]
    \arrow[d, "e_R", two heads]
    \arrow[shiftarr = {xshift=30},->>]{dd}{(!,e)}
    \\
    (Δ^⭑/\bar{e}, D')
    \arrow[d, "{(!,id)}"', two heads]
    \arrow[r, "{(\hat{h}, \hat{f})}", dashed]
    &
    (Σ^⭑/e, D)
    \arrow[d, "{(!, id)}", two heads]
    \\
    (1, D')
    \arrow[r, "{(!,\hat{f})}"', dotted]
    &
    (1, D)
  \end{tikzcd}
\end{equation*}
By the homomorphism theorem, it suffices to show that $\bar{e}_R^⭑(w) = \bar{e}_R^⭑(w')$ implies $e_R^\star(f(w))=e_R^\star(f(w'))$. From the assumption  $\bar{e}_R^⭑(w) = \bar{e}_R^⭑(w')$ it follows that $\bar{e}(ι(w)) = \bar{e}(ι(w'))$ by the definition of $\equiv_{\bar{e}}$, and so
\begin{align*}
ι(e_R^⭑(f^⭑(w))) = e(f(ι(w))) = \hat{f}(\bar{e}(ι(w)))
=  \hat{f}(\bar{e}(ι(w'))) = ι(e_R^⭑(f^⭑(w'))).
\end{align*}
Since $(Σ^⭑/e, D)$ is reduced and $\star$-generated and so ι is injective by \autoref{lem:reduced-vs-embedded}(3), we get $e_R^⭑(f^⭑(w))=e_R^⭑(f^⭑(w'))$, as required.

This proves that $e_R·f$ factors through $\bar{e}_R$ via $(\hat{h}, \hat{f})$ and we are done.
\qed
\end{enumerate}

\section{Details for \autoref{sec:languages}}


\bigskip\noindent\textbf{Proof of \autoref{lem:recognizes-regular}}\\
Recall that a language $L\colon \Sigma^\star\to 2$ is regular iff it is recognizable by a finite  monoid; that is, there exists a finite monoid $M$, a monoid homomorphism $h\colon \Sigma^\star\to M$ and a map $p\colon M\to 2$ such that $L=p\cdot h$. 

Suppose that $L\colon \Sigma^\star\to 2$ is recognized by a finite lattice bimodule $(M, D)$ via $h  \: \f → (M, D) $ and $p \: D → 2$. Then $h^\star$ is a monoid homomorphism that recognizes $L$ via $p·ι$. Since $M$ is a finite monoid, this proves that $L$ is regular.

Conversely, if $L$ is regular, then there exists a finite monoid $M$,
a monoid homomorphism $h \: Σ^⭑ \to M$ and a map $p\colon M\to 2$ with
$L=p\cdot h$. We may assume $h$ to be surjective; if necessary,
replace $h$ by its codomain restriction
$h\colon \Sigma^\star\epito h[\Sigma^\star]$. Thus, the outside in the
diagram below commutes.
    
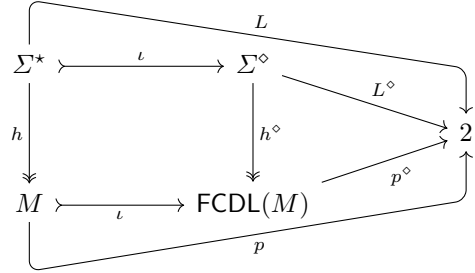
\begin{equation*}
        \begin{tikzcd}[baseline = (M.base), row sep=1.2em]
            Σ^⭑
                \arrow[rr, "ι", tail]
                \arrow[dd, "h"', two heads]
                \arrow[shiftarr = {yshift=25}]{rrrrd}{L}
                & &
            Σ^⋄
                \arrow[dd, "h^\diamond", two heads]
                \arrow[rrd, "L^⋄"] & &  \\
                & & & & 2 \\
            |[alias = M]| M
                \arrow[rr, "ι"', tail]
                \arrow[shiftarr = {yshift=-25}]{rrrru}[swap]{p}
                & &
            \fcdl(M)
                \arrow[rru, "{p^\diamond}"'] & &
              \end{tikzcd}
      \end{equation*}
 In analogy to the proof of \autoref{prop:lbm-free}, the pair $(M, \fcdl(M))$ carries a canonical lattice bimodule structure and the monoid homomorphism $h$ extends to a lattice bimodule homomorphism $(h, h^\diamond) \: \f ↠  (M, \fcdl(M))$. Let $p^\diamond$ to be the extension of $p$ to a $\CDL$-morphism. Then $(h, h^\diamond)$ is a lattice bimodule homomorphism that recognizes $L$ via ${p^\diamond}$. Note that $\fcdl(M)$ is finite because $M$ is finite.\qed

\bigskip\noindent\textbf{Proof of \autoref{cor:reduced-generated-recognizes-regular}}\\
The lattice bimodule $(M,\fcdl(M))$ constructed in the proof of~\autoref{lem:recognizes-regular} has these properties: It is $\star$-generated and $\star$-embedded by definition, and thus reduced by~\autoref{lem:reduced-vs-embedded}.
\qed

\begin{rem}[Languages over \U-quotients]\label{rem:lang-rec-U}
By  \autoref{cor:reduced-generated-recognizes-regular} we know that it suffices to work with ⭑-generated reduced
  lattice bimodules. There is an obvious equivalent formulation of
  language recognition in terms of finite \U-quotients. We therefore
  use the same terminology as for lattice bimodules, i.e.\ say that a
  language $L \: Σ^⭑ → 2$ is \emph{recognized} by a finite \U-quotient
  $e \: Σ^⋄ ↠ D$ via a $\CDL$-morphism $p \: D → 2$ if $L^⋄ =
  p·e$. We denote the set of
  languages recognized by a finite \U-quotient $e$ by \rec{e}. Thus, a surjective lattice bimodule homomorphism $h$ recognizes the languages $L$ iff its corresponding $\U$-quotient $h^\diamond$ recognizes $L$.
\end{rem}

\takeout{
\subsection{Syntactic Lattice Bimodules}
\label{Asec:syntactic-lattice-bimodules}

For any class of automaton or algebras serving as recognizers for a given type of language, it is important to characterize the structure of minimal recognizers. For instance, the minimal monoid recognizing a given language $L \: Σ^⭑ → 2$, a.k.a.~its \emph{syntactic monoid}, is the quotient monoid $e_L\colon \Sigma^\star\epito \Sigma^\star/\mathop{\equiv_L}$ corresponding to the congruence $≡_L$ on $Σ^⭑$ given by
  \begin{align*}
  w ≡_L w' \quad\textrm{iff}\quad ∀x, y ∈ Σ^⭑ \: L(xwy) = L(xw'y).
  \end{align*}
The syntactic monoid is minimal in these sense that the homomorphism $e_L$ recognizes $L$ and factorizes through every other surjective monoid homomorphism $e\colon \Sigma^\star\epito M$ recognizing $L$. Syntactic structures have also been investigated for other types of recognizers for regular languages, including ordered monoids~\cite{pin-1995}, idempotent semirings~\cite{polak2001}, or lattice algebras~\cite{kli-pol}. In this section we derive and examine syntactic structures for lattice bimodules, building on the close relationship between algebraic recognition and unary presentations.

 Recall that language recognition can equivalently be defined via lattice bimodule homomorphisms or via \U-quotients, see \autoref{rem:lang-rec-U}. With the case of syntactic monoids in mind, this suggests the following definition:

  \begin{defn}[Syntactic lattice congruence]
    For a language $L \: Σ^⭑ → 2$ the \emph{syntactic lattice congruence} $\sim_L$ on $Σ^⋄$ is defined by
    \begin{align*}
      V \sim_L W \quad\textrm{iff}\quad [∀x, y ∈ Σ^⭑\: L^⋄(x \tr V \tl y) = L^⋄(x \tr W \tl y)],
    \end{align*}
    or, equivalently, via the unary presentation of \f by
    \begin{align*}
            V \sim_L W \quad\textrm{iff}\quad [∀u ∈ \barU(⋄,⋄) \: L\cdot u(V) = L\cdot u(W)].
    \end{align*}
  \end{defn}

  \begin{lemma}
    The syntactic lattice congruence of a regular language $L$ is a congruence of completely distributive lattices and the resulting CDL quotient $φ_L$ is a \U-quotient.
  \end{lemma}
  \begin{proof}
    The relation is obviously an equivalence relation. Let $I$ be an index set $V_i, W_i ∈ Σ^⋄$ and $V_i \sim_L W_i$ for all $i ∈ I$. We show that the relation is stable under joins, i.e.~$⋁_{i ∈ I} V_i \sim_L ⋁_{i ∈ I} W_i$. Indeed, for all $x, y ∈ Σ^⭑$ we have
    \begin{align*}
      L^⋄(x \tr (⋁_{i ∈ I} V_i) \tl y)
      =&\, L^⋄(⋁_{i ∈ I} (x \tr V_i \tl y) )\\
      =&\, ⋁_{i ∈ I} L^⋄(x \tr V_i \tl y )\\
      =&\, ⋁_{i ∈ I} L^⋄(x \tr W_i \tl y )\\
      =&\, L^⋄(⋁_{i ∈ I} (x \tr W_i \tl y) )\\
      =&\, L^⋄(x \tr (⋁_{i ∈ I} W_i) \tl y)
    \end{align*}
    which proves $⋁_{i ∈ I} V_i \sim_L ⋁_{i ∈ I} W_i$. The case for $⋀$ is analogous and this proves $\sim_L$ a congruence of CDLs.

Next we show the induced CDL quotient $π \: Σ^⋄ ↠ Σ^⋄/\!\!\sim_L$ is a \U-quotient by providing a lifting for all $u ∈ \barU(⋄,⋄)$. All operations in $\barU(⋄,⋄)$ are of the form $u_{v, w} \: W ↦ v \tr W \tl w$ for $v, w ∈ Σ^⭑$. Their liftings $\overline{u}_{v, w}$ needs to satisfy $\pi\cdot u_{v,w}=\overline{u}_{v,w}\cdot \pi$, so the only possible definition is $\overline{u}_{v, w} \: [W]_{\sim_L} ↦ [v \tr W \tl w]_{\sim_L} $. To prove this to be well-defined, suppose that $W \sim_L W'$. Then, for all $x, y ∈ Σ^⭑$,
    \begin{align*}
      L^⋄(x \tr (v \tr W \tl w) \tl y)
      &= L^⋄(xv \tr W \tl wy)\\
      & = L^⋄(xv \tr W' \tl wy)\\
      & = L^⋄(x \tr (v \tr W' \tl w) \tl y),
    \end{align*}
    and so $[v \tr W \tl w]_{\sim_L}=[v \tr W' \tl w]_{\sim_L}$. This proves well-definedness of $\overline{u}_{v, w}$, showing that  $π \: Σ^⋄ ↠ Σ^⋄/\!\!\sim_L$ is a \U-quotient.
  \qed\end{proof}

\newcommand{\synL}[1]{\ensuremath{\mathbf{Syn}_\textrm{#1}(L)}}
We call a \U-quotient $φ \: Σ^⋄ ↠ ~\synL{\U}$ that recognizes a language $L \: Σ^⭑ → 2$ the \emph{syntactic \U-quotient} for $L$ if it is the smallest \U-quotient that recognizes $L$: If $γ \: Σ^⋄ ↠ D$ is another \U-quotient that recognizes $L$ then φ factors through γ:
\begin{equation*}
  \begin{tikzcd}
    Σ^⋄
    \arrow[r, "γ", two heads]
    \arrow[rd, "φ", two heads]
    &
    D
    \arrow[d, "", dashed]
    \\
    &
    \synL{\U}
  \end{tikzcd}
\end{equation*}

  \begin{proposition}
   For a language $L \: Σ^⭑ → 2$, the \U-quotient $φ_L$  induced by the syntactic lattice congruence $\sim_L$ is the syntactic \U-quotient for $L$, so $Σ^⋄/\!\!\sim_L\,= \synL{\U}$.
  \end{proposition}
  \begin{proof}
    First we show $φ_L$ recognizes $L$ via the lattice morphism $ω_L$ given by
    \begin{align*}
ω_L([W]_{\sim_L}) = L^⋄(W)
    \end{align*}
    It is well defined since if $V \sim_L W$ then $∀x, y ∈ Σ^⭑\: L^⋄(x \tr V \tl y) = L^⋄(x \tr W  \tl y)$ and so especially $L^⋄(V) = L^⋄(ε \tr V \tl ε) = L^⋄(ε \tr W \tl ε) = L^⋄(W)$. By definition $ω_L·φ_L = L^⋄$ holds, so $φ_L$ recognizes $L$ via $ω_L$.
    Now let $φ \: Σ^⋄ ↠ D$ be another \U-quotient that recognizes $L$,
    so $L^⋄ = ω·φ$ for some $ω \: D → 2$. We use the homomorphism
    theorem to show that $φ_L$ factors through φ: If $V, W ∈ Σ^⋄$ with
    $φ(V) = φ(W)$ then also $V \sim_L W$, as for all $u\in
    \barU(⋄,⋄)$,
    \[
      \begin{array}{r@{\,}l@{\,}l}
        (L^⋄·u)(V) &= (ω·φ·u)(V) &= (ω·\bar{u}·φ)(V)\\
        &= (ω·\bar{u}·φ)(W) &= (ω·φ·u)(W) = (L^⋄·u)(W),
      \end{array}
    \]
      so $φ_L(V) = φ_L(W)$, and $φ_L$ factors through φ.
    \qed\end{proof}

  In analogy to the syntactic \U-quotient, we also can define the \emph{syntactic lattice bimodule} $s_L \: \f ↠ \synL{LBM}$  to be the smallest reduced lattice bimodule that recognizes $L$: For any reduced lattice bimodule quotient $e$, if $e$ recognizes $L$ then $s_L$ factors through $e$. The results of \autoref{subsec:u-quotients} make the next proof a formality.
  \begin{equation*}
  \begin{tikzcd}
    \f
    \arrow[r, "e", two heads]
    \arrow[rd, "s_L", two heads]
    &
    (M, D)
    \arrow[d, "", dashed]
    \\
    &
    \synL{\U}
  \end{tikzcd}
\end{equation*}

  \begin{corollary}
    The syntactic lattice bimodule is the reduced lattice bimodule quotient induced by the syntactic \U-quotient, that is, $s_L = (φ_L)_R$.
  \end{corollary}
  \begin{proof}
    The proof is just another application of  \autoref{reduced-quotients-props}: $s_L$ is reduced by \ref{reduced-quotients-props}\ref{reduced-quotients-props:2} and it recognizes $L$ as $φ_L$ already recognizes $L$. For any other lattice bimodule quotient that recognizes $L$, by \autoref{rem:gen-quot-u-quot} we get $e^⋄$ is a \U-quotient and as $e^⋄$ recognizes $L$ also $φ_L ≤ e^⋄$. By \ref{reduced-quotients-props}\ref{reduced-quotients-props:2} this proves $s_L = (φ_L)_R ≤ e$, so $s_L$ factorizes through $e$.
  \qed\end{proof}

  Let us take a closer look at the congruence $\equiv_{\phi_L}$ on \f induced by the syntactic lattice bimodule of a language $L$; see~\autoref{not:eqphi}. We know by  \autoref{reduced-quotients-props}.3 that the ⋄-component of \synL{LBM} is equal to \synL{\U}, so the $\diamond$-component of $\equiv_{\phi_L}$ is equal to the syntactic lattice congruence. As for the ⭑-component, we get the following result.

  \begin{lemma}
    The $\star$-component of $≡_{φ_L}$ is the syntactic monoid congruence for $L$.
  \end{lemma}
  \begin{proof}
The set $\barU(⭑, ⋄)$ consists of the unary operations
\[
    x ↦ ι(wxw'),\quad x ↦ wxw' \tr V \tl v,\quad x ↦ v \tr V \tl wxw' \qquad (w, w', v ∈ \Sigma^\star,V\in \Sigma^\diamond).
\]
  Hence, by definition
  \begin{align*}
    &&                                      \phantom{\:}&& w           &≡_{φ_L} w'\\
    \textrm{ iff }&&∀u ∈ \barU(⭑,⋄)                  \: && φ_L(u(w))   &= φ_L(u(w'))\\
    \textrm{ iff }&&∀u ∈ \barU(⭑,⋄)                  \: && u(w)        &\sim_L u(w')\\
    \textrm{ iff }&&∀u ∈ \barU(⭑,⋄), ∀u' ∈ \barU(⋄,⋄)\: && L^⋄(u'(u(w))) &= L^⋄(u'(u(w')))\\
    \textrm{ iff }&&∀u ∈ \barU(⭑,⋄)                  \: && L^⋄(u(w))     &= L^⋄(u(w')),\\
  \end{align*}
since \barU is closed under both composition and identities. Hence, as \barU is given, $w ≡_{φ_L} w'$ iff for all $x, x', y, y' ∈ Σ^⭑$ and $V = ⋁_i⋀_jv_{ij} ∈ Σ^⋄$ the following hold:
  \begin{enumerate}
  \item
    $L^⋄(ι(xwy)) = L^⋄(ι(xw'y))$, that is, $xwy ∈ L \Leftrightarrow xw'y ∈ L$.
  \item
    By substituting $z_{ij} = x'v_{ij}y$ we can compute
    \begin{align*}
      && L^⋄(xwx' \tr V \tl y ) &= L^⋄(xw'x' \tr V \tl y )\\
      \Leftrightarrow&& L^⋄(xwx' \tr ⋁_i⋀_jv_{ij} \tl y ) &= L^⋄(xw'x' \tr ⋁_i⋀_jv_{ij} \tl y )\\
      \Leftrightarrow&& L^⋄( ⋁_i⋀_j xwx'v_{ij}y) &= L^⋄( ⋁_i⋀_j xwx'v_{ij}y)\\
      \Leftrightarrow&& ⋁_i⋀_j L^⋄(xwz_{ij}) &= ⋁_i⋀_j L^⋄(xwz_{ij})\\
    \end{align*}
    and this holds if and only if 1. holds.
  \item
    $L^⋄(x \tr V \tl ywy' ) = L^⋄(x \tr V \tl yw'y' )$ also holds if 1. holds. This is analogous to 2.
  \end{enumerate}
  We see that that two words $w, w' ∈ Σ^⭑$ are identified by the syntactic lattice bimodule congruence if and only if $∀x, y ∈ Σ^⭑\: xwy ∈ L \Leftrightarrow xw'y ∈ L$.
\qed\end{proof}
  This lemma allows us to either interpret $≡_{φ_L}$ as the extension of the syntactic monoid congruence $≡_L$, or to see $≡_L$ as the restriction of $≡_{φ_L}$, depending on where we start.
}


\bigskip\noindent\textbf{Proof of \autoref{thm:iso-thlang-varlang}}
\begin{enumerate}
\item We first prove an auxiliary statement $(⭑)$:
\begin{center}
  If $V_Σ$ is a local basic varietiy of regular languages over the alphabet Σ, every finite subset $S ⊆_f V_Σ$ is contained in a finite local subvariety of $V_Σ$.
\end{center}
To see this, let $S ⊆_f V_Σ$. For any $L\in S$, let $\mathbf{Deriv}(L)$ denote the set of all derivatives of $L$. It is finite since $L$ is regular, and a local subvariety of $V_Σ$ by definition. Since $S$ is finite, the union $F = \bigcup_{L ∈ S}\mathbf{Deriv}(L)$ is again a finite local subvariety and $S ⊆ F ⊆_f V_Σ$.
\item  We show that the two functions given in the statement of the theorem are well-defined, i.e.~map varieties to cotheories and vice versa. For any local basic variety of languages $V_Σ$ the set \[I^{V_Σ}=\{\, F \seq V_\Sigma  \mid F \textrm{ is a finite basic local subvariety of $V_Σ$} \}\] is an ideal of finite local subvarieties of $V_Σ$: it is clearly downwords closed, and it is upwards directed because unions of basic local varieties are basic local varieties. For the same reason the set $\bigcup I_Σ$ is a local basic variety for any ideal $I_\Sigma$ of finite local varieties.

 If $T = (I_Σ)_{Σ ∈ \A}$ is a cotheory of regular languages then the family $V^T = (\bigcup I_Σ)_{Σ ∈ \A}$ is a basic variety of regular languages: For every monoid homomorphism $g\colon \Delta^\star\to \Sigma^\star$  the function $g^{-1}$ restricts locally to some $F' ∈ I_Δ$ for all $F ∈ I_Σ$, so
\begin{align*}
g^{-1}[\bigcup I_Σ] = g^{-1}[\bigcup_{F ∈ I_Σ} F] = \bigcup_{F ∈ I_Σ} g^{-1}[F] ⊆ \bigcup_{F ∈ I_Σ} F' ⊆ \bigcup I_Δ
\end{align*}
and thus $g^{-1}L ∈ \bigcup I_Δ$ for all languages $L ∈ \bigcup I_Σ$.

Conversely, let $V = (V_Σ)_{Σ ∈ \A}$ be a basic variety of regular languages. To show $T = (I^{V_Σ})_{Σ ∈ \A}$ a cotheory of regular languages, let $g  \: \Delta^\star\to \Sigma^\star$, and take some $F ∈ I^{V_Σ}$. Since $F$ is finite, we have that $g^{-1}[F]$ is a finite basic local subvariety of $V_\Delta$; it is closed under derivatives because $v^{-1}(g^{-1}L)w^{-1} = g(v)^{-1}Lg(w)^{-1}$ for all $v,w\in \Delta^\star$. Thus, $g^{-1}[F]$ lies in $V_\Delta$. 

\item It remains to prove that the two constructions are mutually inverse. First we show that $V_Σ = \bigcup I^{V_Σ}$ for any local basic variety $V$. By $(⭑)$, for every $L ∈ V_Σ$ we have $L ∈ F ⊆ V_Σ$ for some finite local subvariety $F$, so $L ∈ F ⊆ \bigcup I^{V_Σ}$. For the other direction we have $\bigcup I^{V_Σ} ⊆ \bigcup \Pow(V_Σ) = V_Σ$.\\
Next we have to show for any basic cotheory $I=(I_\Sigma)_{\Sigma\in \Setf}$ that $I_\Sigma = I^{V_Σ}$ with $V_Σ = \bigcup I_\Sigma$. This time the inclusion $I_\Sigma ⊆ I^{V_Σ}$ is clear. Now let $F\in I_\Sigma^{V_\Sigma}$. Then each $L ∈ F$ is element of some  $F_L\in I_\Sigma$. Since $F ⊆ \bigcup_{L\in F} F_L ∈ I$ and $I_\Sigma$ is downwards closed, this proves $F ∈ I$.

Both assigments are clearly order-preserving.\qed
\end{enumerate}

\section{Details for \autoref{sec:variety-theorem}}
\label{Asec:variety-theorem}
We first provide some details about the duality between algebraic completely distributive lattices and posets. Let us start by recalling some standard terminology from order theory~\cite{dav-pri}.
A subset $D ⊆ P$ of a poset $P$ is a \emph{down-set} if $p\in D$ and $p' ≤ p$ implies $p' ∈  D$. For $p ∈ P$ the down-set $\ds{p} = \{a ∈ P \pipe a ≤ p\}$ is called the \emph{principal down-set} of $p$. Since intersections and unions of down-sets are again down-sets we see that the set $\mathcal{D}(P)$ of down-sets of $P$ forms a completely distributive lattice. An element $c$ of a complete lattice $D$ is \emph{compact} if whenever $c ≤ ⋁S$ then $c ≤ \bigvee F$ for some finite subset $F ⊆_f S$, and \emph{join-prime} if $c ≤ ⋁S$ implies $c ≤ s$ for some $s ∈ S$. We denote the sets of compact and join-prime elementsif of a lattice $D$ with $K(D)$ and $J_p(D)$, respectively. A complete lattice $D$ is \emph{algebraic} if every element is the join of all compact elements below it, that is, $d = ⋁(\ds{d} \cap K(D))$ for all $d\in D$. Algebraic CDLs form a full subcategory of \CDL denoted \AlgDL. For any poset $P$ the lattice $\mathcal{D}(P)$ is algebraic; its join-primes are the principal downsets $\ds{p}$ ($p\in P$), and its compact elements are finitely generated down-sets (i.e. finite unions of principal down-sets).

\begin{proposition}
  The category \AlgDL is dually equivalent to the category $\POS$ of posets and monotone maps, witnessed by the equivalence functor
\[ \mathcal{D}\colon \POS \xrightarrow{\simeq} \AlgCDL^\op \]
that maps a poset $P$ to the lattice $\mathcal{D}(P)$ of down-sets, and a monotone map $f\colon P\to Q$ to the CDL morphism $\mathcal{D}(f)=f^{-1}\colon \mathcal{D}(Q)\to \mathcal{D}(P)$.
\end{proposition}
We provide a proof of this well-known duality for the convenience of the reader.

\begin{proof}
  By~\cite[Thm.~10.29]{dav-pri} a CDL is algebraic iff it is
  isomorphic to $\mathcal{D}(P)$ for some poset $P$, which implies
  that $\mathcal{D}$ is isomorphism-dense. To show that $\mathcal{D}$ is an
  equivalence functor, it remains to prove that it is full and faithful.

  To see that $\mathcal{D}$ is faithful, let $f, g \: P → Q$ with
  $\mathcal{D}(f) = \mathcal{D}(g)$. Then for all $q\in Q$ we have
  we have $\mathcal{D}(f)(\ds{q}) =
  \mathcal{D}(g)(\ds{q})$, that is, for all $p\in P$, 
  \begin{align*}
    f(p)\leq q \quad \text{iff} \quad g(p)\leq  q. 
  \end{align*}
  Substituting $f(p)$ and $g(p)$ for $q$ shows that
  $f(p) ≤ g(p)$ and $g(p) ≤ f(p)$ for all $p ∈ P$, thus proving $f=g$. This shows that $\mathcal{D}$ is faithful.

  Finally, we show that $\mathcal{D}$ is full.  Let
  $g \: \mathcal{D}(Q) → \mathcal{D}(P)$ be a CDL morphism. For each
  $p ∈ P$ let $D_p$ denote the least down-set of $Q$ whose $g$-image
  contains $p$:
  \[
    D_p = \bigcap \{D ∈ \mathcal{D}(Q) \pipe p ∈ g(D)\} ∈
    \mathcal{D}(Q).
  \]
Note that $p\in g(D_p)$ because $g$ preserves intersections.
The set $D_p$ is join-prime since
  if $D_p ⊆ \bigcup_i K_i$ then $p ∈ g(K_j)$ for some $j$, and so
  $D_p ⊆ K_j$ by the minimality of $D_p$. The join-prime elements of
  $\mathcal{D}(Q)$ are precisely the principal down-sets of $Q$, hence
  $D_p =\,\,\ds{f(p)}$ for some unique $f(p) ∈ Q$. This defines a
  monotone map $f \: P → Q$ via $p ↦ f(p)$ for which we prove
  $f^{-1} = g$. If $D ∈ \mathcal{D}(Q)$, then
  \begin{align*}
    p ∈ f^{-1}(D) \Leftrightarrow f(p) ∈ D \Leftrightarrow\,\ds{f(p)} ⊆ D \Leftrightarrow D_p ⊆ D \Leftrightarrow \bigcap_{p ∈ g(K)}K ⊆ D \Leftrightarrow p ∈ g(D)
  \end{align*}
  which gives $f^{-1}(D) = g(D)$ and so  $g=f^{-1}=\mathcal{D}(f)$, proving that $\mathcal{D}$ is full.
\qed\end{proof}

\begin{rem}
\begin{enumerate}
\item We may identify $\mathcal{D}(P)$ with the lattice $\POS(P, 2)$ of monotone functions into the two-chain $2=\{0<1\}$ via
  \begin{align*}
    D &↦ χ_{D^c},\\
     f^{-1}[0] &\mapsfrom f,
  \end{align*}
  where $χ_{D^c} \: P → 2$ denotes the characteristic function of the complement of $D$. Thus, up to natural isomorphism, the equivalence $\mathcal{D}$ is given by the hom-functor \[\POS(-, 2) \: \POS \xrightarrow{\simeq} \AlgDL^\op.\] 
\item It is also instructive to see how the duality operates in the other direction: From~\cite[Theorem 10.29]{dav-pri} we know that for a completely distributive algebraic lattice $D$ the poset $P$ with $\mathcal{D}(P) \cong D$ is isomorphic to $J_p(D)$, the poset of join-prime elements of $D$, but equipped with the \emph{dual order} of $D$: for $p, q ∈ J_p(D)$ we have $p ≤_{J_p(D)} q$ iff $q ≤_D p$. Note that we have the isomorphism $J_p(D) \cong \AlgDL(D, 2)$ by identifying an element of $d\in J_p(D)$ with the morphism $f\colon D\to 2$  sending $d'\in D$ to $1$ iff $d\leq d'$. The order $\leq_{J_p(D)}$ is the order induced by the pointwise ordering on $\AlgDL(D, 2)$. Thus, the inverse equivalence of $\mathcal{D}$ is, up to natual isomorphism, the hom-functor \[\AlgCDL(-,2)\colon \AlgCDL^\op \xrightarrow{\simeq} \POS.\]
\end{enumerate}
\end{rem}

\newcommand{\PLang}[1][Σ]{\ensuremath{\Pow({#1}^⭑)}\xspace}

\begin{rem}
To make use of the duality $\AlgDL^\op \cong \POS$ in our setting, let us note that the concept of a $\U$-quotient $e\colon \Sigma^\diamond \epito D$ actually ``lives'' in the full subcategory $\AlgDL$ of $\CDL$ since it involves only free or finite CDLs. Every free CDL $\Sigma^\diamond$ is algebraic since the elements of the form $\bigwedge_{i\in I} w_i$ ($w_i\in \Sigma^\star$) are join-prime. Moreover, every finite lattice $D$ is algebraic because every element $d\in D$ is compact. Furthermore, the factorization system of \CDL restricts to \AlgDL, and we can thus safely adopt our concepts of quotients into \AlgDL:
\end{rem}

\begin{lemma}
  \AlgDL inherits the factorization system of surjective and injective morphisms from \CDL.
\end{lemma}
\begin{proof}
  For the proof we use that a lattice is an algebraic CDL if and only if it is isomorphic to a complete lattice of sets~\cite[Theorem 10.29]{dav-pri}. So let $h \: L → M$ be a function between complete lattices of sets that preserves arbitrary unions and intersections. In \CDL the morphism $h$ factorizes into
  \begin{equation*}
    L \stackrel{p}{↠} K \stackrel{i}{↣} M
  \end{equation*}
  with $p$ surjective and $i$ injective. Since $K$ is a complete sublattice of $M$ and thus also isomorphic to a lattice of sets, we see that $K$ is algebraic. Thus, $h=i\cdot p$ is a factorization of $h$ in $\AlgCDL$.
\qed\end{proof}

\begin{rem}
Note that quotients (i.e. surjective homomorphisms) in $\AlgCDL$ dualize to subposets (i.e. maps $m$ satisfying $x\leq y$ iff $m(x)\leq m(y)$) in $\POS$. This follows immediately from the definition of the dual equivalence, but also from the fact that quotients in $\AlgCDL$ correspond to strong epimorphisms and subposets correspond to strong monomorphisms in $\POS$. Using this duality, $\U$-quotients in $\AlgCDL$ admit a natural dual interpretation in terms of the languages recognized by them (cf.\ \autoref{rem:lang-rec-U}):
\begin{enumerate}
\item
If we start with a finite \U-quotient $e \colon Σ^⋄ ↠ D$ in \AlgDL, it dualizes to the embedding of a finite subposet
\begin{align*}
J_p(e) \colon J_p(D) ↣ J_p(Σ^⋄).
\end{align*}
The join-primes of $Σ^⋄$ are given by $J_p(Σ^⋄) \cong \AlgDL(Σ^⋄, 2) \cong \Set(Σ^⭑, 2) \cong \PLang$, so we may regard $J_p(e)$ as a subobject $J_p(D) ↣ \PLang$, i.e.~a set of languages. Now let $k \: J_p(D) \cong \AlgDL(D, 2) → \rec{e}$ be the bijection given by $p ↦ p·e$. Then, using that $J_p(e)$ is given by precomposition with e, we see that the map $k$ makes the following triangle commute:
\begin{equation*}
  \begin{tikzcd}
    &
    \PLang
    &
    \\
    \AlgDL(D, 2)
    \arrow[ur, "J_p(e)", tail]
    \arrow[rr, "k"', "\cong"]
    &
    &
    \rec{e}
    \arrow[ul, "ι"', tail]
  \end{tikzcd}
\end{equation*}
Therefore $\AlgDL(D, 2)$ and $\rec{e}$ are isomorphic subposets of $\Pow(\Sigma^\star)$. Note that all elements of $\rec{e}$ are regular languages by \autoref{lem:recognizes-regular}. Since e is a \U-quotient, for every $u ∈ \barU(⋄,⋄)$ there exists a lifting:
\begin{equation}
  \begin{tikzcd}
    Σ^⋄
    \arrow[r, "u"]
    \arrow[d, "e"', two heads]
    &
    Σ^⋄
    \arrow[d, "e", two heads]
    \\
    D
    \arrow[r, "", dashed]
    &
    D
  \end{tikzcd}
\end{equation}
Dualizing this diagram yields
\begin{equation}
  \begin{tikzcd}
    \PLang
    \arrow[r, "u^{-1}"]
    &
    \PLang
    \\
    \arrow[u, "", hook]
    \arrow[r, "", dashed]
    \rec{e}
    &
    \arrow[u, "", hook]
    \rec{e}
  \end{tikzcd}
\end{equation}
indicating that for all $v, w ∈ Σ^⭑$ the word derivation function $L ↦ v^{-1}Lw^{-1}$ restricts to the subset \rec{e}. In other words, \rec{e} is a finite local basic variety of languages.

\item Conversely, if we start out with a finite local basic variety of languages $i\:V_Σ \hookrightarrow \reg{Σ} \hookrightarrow \PLang$ then $V_Σ$ is a subobject of $\PLang$ in \POS. Its dual $\mathcal{D}(i)$ is therefore a quotient of $Σ^⋄$ in \AlgDL. Since $V_Σ$ is closed under all word derivatives $L ↦ v^{-1}Lw^{-1}$ represented by elements $u\in \barU(⋄,⋄)$, the map $u^{-1}$ on $\Pow(\Sigma^\star)$ restricts to $V_\Sigma$, i.e.\ we have the commutative diagrams
 \begin{equation*}
   \begin{tikzcd}
    \arrow[r, "u^{-1}"]
    \PLang
    &
    \PLang
    \\
    \arrow[u, "i", hook]
    \arrow[r, "", dashed]
    V_Σ
    &
    \arrow[u, "i"', hook]
    V_Σ
  \end{tikzcd}
\end{equation*}
Thus, dually, $\mathcal{D}(i)$ is a quotient in \AlgDL such that every $u ∈ \U$ has a lifting:
\begin{equation*}
  \begin{tikzcd}
    Σ^⋄
    \arrow[r, "u"]
    \arrow[d, "\mathcal{D}(i)"', two heads]
    &
    Σ^⋄
    \arrow[d, "\mathcal{D}(i)", two heads]
    \\
    \mathcal{D}(V_Σ)
    \arrow[r, "", dashed]
    &
    \mathcal{D}(V_Σ)
  \end{tikzcd}
\end{equation*}
This proves that $\mathcal{D}(i)$ is a \U-quotient for any finite local basic subvariety $i \: V_Σ \hookrightarrow \PLang$. More specifically, $\mathcal{D}(i)$ is the \U-quotient recognizing precisely the languages in $V_Σ$; we see this since any $L ∈ V_Σ$ is representable by the triangle
\[
\begin{tikzcd}[cramped, sep=small]
  \PLang
  &
  &
  \\
  &
  &
  1
  \arrow[llu, "L"']
  \arrow[lld, "L"]
  \\
  V_Σ
  \arrow[uu, "i", hook]
\end{tikzcd}
\qquad\text{that dualizes to}\qquad
\begin{tikzcd}[cramped, sep=small]
  Σ^⋄
  \arrow[dd, "{\POS(i, 2)}"', two heads]
  \arrow[rrd, "L"]
  &
  &
  \\
  &
  &
  2 \cong \mathcal{D}(1)
  \\
  \mathcal{D}(V_Σ)
  \arrow[rru, "{\POS(L, 2)}"']
\end{tikzcd}
\]
proving that $\mathcal{D}(V_Σ)$ recognizes $L$. Conversely, any $L$ recognized by $\mathcal{D}(V_Σ)$ dualizes to some element of $V_Σ$ if we start with the triangle in \AlgDL.
\end{enumerate}
\end{rem}
We have thus established the following result:
\begin{proposition}\label{lem:finiteuq-vs-finitelbv}
The lattice of finite $\U$-quotients of $\Sigma^\diamond$ is isomorphic to the lattice of finite local basic varieties over $\Sigma$. The isomorphism is given by 
\[ e\quad\mapsto\quad (\,\rec{e} \monoto \Pow(\Sigma^\star)\,).\]
\end{proposition}
This isomorphism easily extends to the level of ideals:
\begin{corollary}[Duality between local varieties]
For each $Σ ∈ \A$ the lattice of local pseudovarieties of \U-quotients over $\Sigma$ is isomorphic to the lattice of ideals $\mathcal{I}_Σ$ of finite local basic varieties over Σ. The isomorphism is given by
\[ \locT \quad\mapsto\quad \{\,\rec{e}\;:\; e\in \locT \, \}. \]
\end{corollary}

\begin{rem}
  \sloppypar
As the final step, we observe that the above local correspondence extends to a global one between theories of $\U$-quotients and basic cotheories of regular languages. Suppose that $\Th = (\Th_Σ)_{Σ ∈ \A}$ is a theory of \U-quotients. Thus, for all lattice bimodule homomorphisms $h \colon \f[Δ] → \f[Σ]$ and every $e ∈ \Th_Σ$ there exists a lifting of $e·h^⋄$ through $\locT[Δ]$:
\begin{equation*}
\begin{tikzcd}
  Δ^⋄
  \arrow[d, "\bar{e}"', two heads]
  \arrow[r, "h^⋄"]
  &
  Σ^⋄
  \arrow[d, "e", two heads]
  \\
  D'
  \arrow[r, "", dashed]
  &
  D
\end{tikzcd}
\end{equation*}
Thus, letting $g=h^\star$ denote the monoid morphism in the first component of $h$, the dual diagram in \POS then precisely states that the corresponding family of ideals of finite basic local varieties is closed under preimages of $g$, and vice versa.

\begin{equation*}
  \begin{tikzcd}
    \PLang[Δ]
    &
    \PLang[Σ]
    \arrow[l, "g^{-1}"']
    \\
    \rec{\overline{e}}
    \arrow[u, hook]
    &
    \rec{e}
    \arrow[u, hook]
    \arrow[l, "", dashed]
\end{tikzcd}
\end{equation*}
We have thus established the following result:
\end{rem}

\begin{proposition}[Duality between theories and cotheories]\label{obs:duality}
The lattice of theories of \U-quotients is isomorphic to the lattice of basic cotheories of regular languages. The isomorphism is given by
\[ \Th \quad\mapsto\quad  T=(I_\Sigma)_{\Sigma\in\Set_f} \text{ with } I_\Sigma= \{\,\rec{e}\;:\; e\in\locT\,\}.\]
\end{proposition}

\bigskip\noindent\textbf{Proof of \autoref{variety-theorem}}\\
We simply compose all the previously established lattice isomorphisms:
\begin{align*}
  \phantom{\cong}
  &\textrm{ Pseudovarieties of lattice bimodules}
  \\
  \cong
  &\textrm{ Theories of lattice bimodules}
  & \text{(\autoref{thm:iso-lbm-rt})}
  \\
  \cong
  &\textrm{ Theories of \U-quotients}
  &\text{(\autoref{prop:iso-thUquot-thLBM})}
  \\
  \cong
  &\textrm{ Basic cotheories of regular languages}
  &\text{(\autoref{obs:duality})}
  \\
  \cong
  &\textrm{ Basic varieties of regular languages}
  &\text{(\autoref{thm:iso-thlang-varlang})}\tag*{\qed}
  \end{align*}

\takeout{
\subsection{Kondracs-Watrous quantum finite automata}
\label{Asec:kwqfa}

\begin{defn}
  A \emph{Kondacs-Watrous quantum finite automaton (KWQFA)} is given
  by a tuple
  $M = (Q, \Sigma, \{A_{\sigma} \pipe \sigma \in \tilde{\Sigma} \},
  q_0, Q_\mathsf{acc}, Q_\mathsf{rej}, Q_\mathsf{non})$ with $Q$ a
  finite set of (basis) states, $\Sigma$ the input alphabet not
  containing the end markers $\kappa$ and $\$$, for each symbol
  $\sigma \in \tilde{\Sigma}$ a unary transformation
  $A_\sigma$, an initial state $q_0$ and a partition of $Q =
  Q_\mathsf{acc} \dot{\cup} \;Q_\mathsf{rej}\; \dot{\cup}\S;
  Q_\mathsf{non}$ into accepting, rejecting and non-halting states,
  respectively. A state $|\psi\rangle$ of
  $M$ is an element on the unit sphere of the finite
  $|Q|$-dimensional complex Hilbert space
  $\mathcal{H}_{|Q|}$. For each $q \in
  Q$ we denote its corresponding basis state by
  $|q\rangle$, thus
  $|\psi\rangle$ is given as a linearcombination $|\psi\rangle =
  \sum_{q \in Q} \alpha_q|q\rangle$ with $\alpha_q \in
  \mathds{C}$ satisfying $\sum_{q \in Q}\|\alpha_q\| =
  1$. The initial state is given by $|q_0\rangle$. An input $w \in
  \Sigma^\star$ is processed by first adding the left
  ($\kappa$) and right ($\$$) input markers and then applying for
  every symbol $w_i$ in $\tilde{w} = \kappa w \$$ the corresponding
  transformation
  $A_{w_{i}}$ and then perform a measurement whether the state is in
  $Q_\mathsf{acc}$, $Q_\mathsf{rej}$ or
  $Q_\mathsf{non}$ and accordingly react by halting and accept, halting and
  reject or by continuing with processing the rest of the word,
  respectively.
\end{defn}

More specifically,  if a QFA is processing a word and has current state
\[ \ket{\psi} = \sum_{q \in Q_{acc}} \alpha_{q} \ket{q} + \sum_{q' \in Q_{rej}} \beta_{q'}\ket{q'} + \sum_{q'' \in Q_{non}} \gamma_{q''} \ket{q''} \]
while reading the $i$-th symbol $w_{i}$ of $w$ then it accepts after the next measurement with probability $\sum_{q \in Q_{acc}} \|\alpha_{q}\|$, rejects with probability $\sum_{q' \in Q_{rej}} \|\beta_{q'}\|$ and continues processing with probability $\sum_{q'' \in Q_{non}} \|\gamma_{q''}\|$.

There are two main approaches to define language recognition for QFA: Language recognition with \emph{unbounded} and with \emph{bounded} error. Here, we only focus on the second notion, as in the unbounded case KWQFA are equivalent in power with stochastic automata~\cite{yak-say} and thus recognize non-regular languages.

\begin{defn}
  A KWQFA $M$ recognizes a languages $L \subset \Sigma^{\star}$ with \emph{error~bound~$\epsilon$} ($0 \leq \epsilon < 1/2$) if it accepts every word in $L$ with probability greater or equal to $1 - \epsilon$ and accepts every word not in $L$ with probability smaller or equal to $\epsilon$. The class of languages recoginized by KWQFA is denoted $\RMM$.
\end{defn}

\subsection{Reversible automata}
\label{Asec:reversible-automata}

In~\cite{gol-pin} and Pin study reversible finite automata (RFA) as a special case of KWQFA. It is shown that RFAs are equivalent to a class of finite automata called classical reversible finite automata (CRFA). Furthermore, the class of languages $\mathbf{K}$ accepted by CRFA is closed under word derivatives and preimages of monoid homorphisms between free monoids and therefore constitutes a basic variety of regular languages. As $\mathbf{K}$ is closed only under complement, but neither under union nor under intersection it is not an instance of any known variety theorem to the best of our knowledge.
\begin{corollary}
The class of finite reduced lattice bimodules recognizing the same languages as RFAs is a pseudovariety of lattice bimodules.
\end{corollary}

\subsection{Injective automata}
\label{sec:injective-automata}

Injective finite automata (IFA) constitute a special case of CRFA, and are also introduced in~\cite{kli-pol}. An IFA is a CRFA that has at most one absorbing state. IFAs can thus be split into two classes; depending on whether this state is final (IFA-A) or nonfinal (IFA-R). Denoting the classes of languages recognized by IFA-A and IFA-R with $\mathsf{L}$ and $\mathbf{L^c}$, respectively, it is proven in~\cite{gol-pin} that both $\mathsf{L}$ and $\mathbf{L}^\mathsf{c}$ are closed under derivatives and under preimages of morphisms between free monoids, but it is also mentioned that even though $\mathsf{L}$ ($\mathbf{L}^\mathsf{c}$) is closed under finite union (intersection), it does not form a disjunctive (conjunctive) variety as defined in~\cite{polak2001}. So neither $\mathsf{L}$ nor $\mathbf{L^c}$ are known instances of variety theorems until now, but both $\mathsf{L}$ and $\mathbf{L}^\mathsf{c}$ form basic varieties of languages. We therefore get the following result:
\begin{corollary}
The class of finite reduced lattice bimodules recognizing languages from $\mathbf{L}$ $(\mathbf{L^c})$ is a pseudovariety of lattice bimodules.
\end{corollary}
}

\end{document}


%% file: unicode.tex
\usepackage{newunicodechar}

\renewcommand{\epsilon}{\varepsilon}

\newcommand{\monoto}{\hookrightarrow}
\newcommand{\epito}{\twoheadrightarrow}

\newcommand{\barU}[0]{\ensuremath{\overline{\mathbb{U}}_Σ}\xspace}
\newcommand{\mcpower}[2]{\ensuremath{\mathcal{#1}^\mathcal{#2}}\xspace}
\newcommand{\locT}[1][Σ]{\ensuremath{ \mathcal{T}_{#1} }\xspace}
\newcommand{\T}[0]{\ensuremath{ (\locT)_{Σ ∈ \A}}\xspace}

\newcommand{\rec}[1]{\ensuremath{\mathbf{Rec}(#1)}}
\newcommand{\reg}[1]{\ensuremath{\mathbf{Reg}_{#1}}}

\renewcommand{\H}{\mathcal{H}}

\newcommand{\COMMENT}[1]{}
\renewcommand{\:}[0]{\colon}

\newcommand{\RMM}{\mathbf{RMM}}
\newcommand{\Pow}{{\mathcal{P}}}
\newcommand{\C}{\ensuremath{\mathscr{C}}\xspace}

\newcommand{\fm}[1] {
  \ensuremath{{#1}^{\star}}\xspace
}

\newcommand{\fcdl}{\mathop{\mathsf{FCDL}}}
\newcommand{\flb}[2] {
  \ensuremath{\fcdl(\fm{#1} + \fm{#1} \times #2 \times \fm{#1})}\xspace
}

\newcommand{\tl}{
    \ensuremath{\triangleleft}\xspace
}
\newcommand{\tr}{
    \ensuremath{\triangleright}\xspace
}

\newcommand{\pipe}{\mathrel{|}}

\newcommand{\f}[1][Σ]{\ensuremath{(\fm{#1}, {#1}^⋄)}\xspace}
\newcommand{\U}{\ensuremath{\mathbb{U}}\xspace} 
\tikzset{
    subs/.style={anchor=south, rotate=90, inner sep=.5mm}
}

\newcommand{\V}{{\mathcal{V}}}
\newcommand{\Th}{{\mathcal{T}}}

\newcommand{\seq}{\subseteq}
\newcommand{\op}{\mathsf{op}}
\newcommand{\Setf}{\ensuremath{\cat{Set}_{\mathsf{f}}}\xspace}
\newcommand{\LBM}{\ensuremath{\cat{LBM}}\xspace}

\newcommand{\POS}{\ensuremath{\cat{Pos}}\xspace}
\newcommand{\CDL}{\ensuremath{\cat{CDL}}\xspace}
\newcommand{\AlgDL}{\ensuremath{\cat{AlgCDL}}\xspace}
\newcommand{\AlgCDL}{\ensuremath{\cat{AlgCDL}}\xspace}
\newcommand{\cat}[1]{\ensuremath{\mathbf{#1}}\xspace}

\newcommand{\A}{\ensuremath{\Setf}\xspace}

\newunicodechar{×}{\ensuremath{\times}}
\newunicodechar{ø}{\ensuremath{\emptyset}}
\newunicodechar{≤}{\ensuremath{\le}}
\newunicodechar{≥}{\ensuremath{\ge}}
\newunicodechar{∈}{\ensuremath{\in}}
\newunicodechar{∅}{\ensuremath{\emptyset}}
\newunicodechar{⊆}{\ensuremath{\subseteq}}
\newunicodechar{⊗}{\ensuremath{\otimes}}

\newunicodechar{→}{\ensuremath{\to}}
\newunicodechar{⇒}{\ensuremath{\Rightarrow}}
\newunicodechar{↓}{\ensuremath{\downarrow}}
\newunicodechar{↑}{\ensuremath{\uparrow}}
\newunicodechar{↠}{\ensuremath{\twoheadrightarrow}}
\newunicodechar{↣}{\ensuremath{\rightarrowtail}}
\newunicodechar{↦}{\ensuremath{\mapsto}}

\newunicodechar{∧}{\ensuremath{\land}}
\newunicodechar{∨}{\ensuremath{\lor}}
\newunicodechar{⋀}{\ensuremath{\bigwedge}}
\newunicodechar{⋁}{\ensuremath{\bigvee}}
\newunicodechar{∀}{\ensuremath{\forall}}
\newunicodechar{∃}{\ensuremath{\exists}}
\newunicodechar{⊤}{\ensuremath{\top}}
\newunicodechar{⊥}{\ensuremath{\bot}}
\newunicodechar{¬}{\ensuremath{\neg}}

\newunicodechar{∏}{\ensuremath{\prod}}
\newunicodechar{≡}{\ensuremath{\equiv}}
\newunicodechar{≠}{\ensuremath{\neq}}
\newunicodechar{·}{\ensuremath{\cdot}}
\newunicodechar{⋄}{\ensuremath{\diamond}}
\newunicodechar{⭑}{\ensuremath{\star}}
\newunicodechar{⟨}{\ensuremath{\langle}}
\newunicodechar{⟩}{\ensuremath{\rangle}}

\newunicodechar{Σ}{\ensuremath{\Sigma}}
\newunicodechar{η}{\ensuremath{\eta}}
\newunicodechar{λ}{\ensuremath{\lambda}}
\newunicodechar{ε}{\ensuremath{\epsilon}}
\newunicodechar{μ}{\ensuremath{\mu}}
\newunicodechar{Δ}{\ensuremath{\Delta}}
\newunicodechar{φ}{\ensuremath{\phi}}
\newunicodechar{π}{\ensuremath{\pi}}
\newunicodechar{Π}{\ensuremath{\Pi}}
\newunicodechar{ι}{\ensuremath{\iota}}
\newunicodechar{γ}{\ensuremath{\gamma}}
\newunicodechar{Γ}{\ensuremath{\Gamma}}
\newunicodechar{Ψ}{\ensuremath{\Psi}}
\newunicodechar{ψ}{\ensuremath{\psi}}
\newunicodechar{Φ}{\ensuremath{\Phi}}
\newunicodechar{χ}{\ensuremath{\chi}}
\newunicodechar{Θ}{\ensuremath{\Theta}}
\newunicodechar{ω}{\ensuremath{\omega}}
\newunicodechar{Ω}{\ensuremath{\Omega}}
\newunicodechar{α}{\ensuremath{\alpha}}
\newunicodechar{β}{\ensuremath{\beta}}